\documentclass[12pt]{iopart}
\usepackage{iopams,tikz}
\graphicspath{{./pictures/}}
\begin{document}
	
\title[]{Lie sphere-geometry in lattice cosmology}
\author{Michael Fennen$^{1}$ and Domenico Giulini$^{1,2}$}
\address{$^1$ Center for Applied Space Technology and Microgravity, 
University of Bremen, Germany}
\address{$^2$ Institute for Theoretical Physics, 
Leibniz University of Hannover, Germany}

\ead{giulini@itp.uni-hannover.de}

\begin{abstract}
In this paper we propose to use 
\emph{Lie sphere-geometry} as a new tool to 
systematically construct time-symmetric 
initial data for a wide variety of generalised 
black-hole configurations in lattice cosmology. 
These configurations are iteratively 
constructed analytically and may have any 
degree of geometric irregularity. We show 
that for negligible amounts of dust these
solutions are similar to the swiss-cheese 
models at the moment of maximal expansion. 
As Lie sphere-geometry has so far not 
received much attention in cosmology, we 
will devote a large part of this paper to 
explain its geometric background in a 
language familiar to general relativists.  
\end{abstract}



\maketitle


\section{Introduction}
In their seminal paper \cite{LindquistWheeler} of 1957, 
Richard Lindquist and John Wheeler introduced the idea 
to approximate the global dynamics of homogeneous and 
isotropic cosmological models by lattice-like 
configurations of vacuum Schwarzschild geometries. 
Approximate homogeneity and isotropy was translated 
into the requirement that this lattice should be a 
regular one, such that each lattice site is equally 
distant to its nearest neighbours. Hence, approximating 
a round 3-sphere, which for the moment we think of as
embedded into euclidean $\mathbb{R}^4$, this implies 
that the lattice sites are given by the vertices of 
inscribed 4-dimensional regular convex polytopes 
(platonic solids), of which there are 6 in 4 
dimensions, corresponding to  $N=5, 8, 16, 24, 120$ 
and $600$ vertices. 

In order to avoid confusion, the method of lattice 
cosmology has to be clearly distinguished from the 
related but different so-called ``swiss-cheese'' 
models, which we shall briefly describe  and which also 
play some role in this paper. The swiss-cheese models 
are constructed from the homogeneous and isotropic 
models in standard dust-matter cosmology by introducing 
local inhomogeneities as follows: replace the 
spherically-symmetric and locally homogeneous geometry 
in a neighbourhood of a vertex (the method works for any 
point, but in order to compare it with lattice cosmology 
we stick to the vertices) by the spherically-symmetric 
and locally inhomogeneous vacuum Schwarzschild geometry 
with appropriate matching conditions at the boundary 
to the dust-filled complement. The matching conditions 
require the metric to be continuously differentiable 
across the boundary and essentially impose the condition 
that the mass of the black hole equals that of the 
removed dust (they must be strictly equal in terms of the 
Misner-Sharp mass; compare \cite{CarreraGiulini}).
This works for any sizes of balls centred around each 
vertex, as long as the collection of balls have no 
pairwise intersections. Outside the balls the dust 
is still present and the local geometry is still that of 
the round 3-sphere (in case of positive curvature, 
to which we restrict attention here). As already 
stated, inside the balls the geometry is strictly 
spherically symmetric, even though the distribution of 
black holes around them on neighbouring vertices is 
only approximately so. This is because the remaining dust 
just enforces this symmetry by construction. It should be 
intuitively obvious why these are referred to as 
``swiss-cheese'' models. 

In contrast, in lattice cosmology, \emph{all} the dust is 
replaced by a number of black holes, none of which 
will now give rise to a strictly spherically symmetric 
geometry in its neighbourhood. Approximate spherical 
symmetry will be improved by increasing the number of 
black holes, i.e. the number of vertices, but never attained
exactly. There is now no matter present whatsoever and 
all gravitating masses are concentrated in black holes. 
Hence the evolution equations are pure vacuum. 

Now, the central ideas behind lattice cosmology is 
that \emph{as regards certain aspects of the overall 
gravitational dynamics, we may replace all matter by 
an appropriate but fictitious distributions of black 
holes}. The hope connected with this strategy is to 
gain reliable analytical insight into various 
aspects of global gravitational dynamics in cosmology, 
like, e.g., the back-reaction or the fitting 
problem~\cite{Clarkson.EtAl:2011a}. This hope 
rests on the fact that now we are dealing with 
the \emph{vacuum} Einstein equations and its associated initial-value problem, the analytic
treatment of which,
albeit still complicated, is considerably simpler 
than that of the coupled Einstein-matter equations 
for realistic models of matter. In addition, for special 
classes of initial data, the constraint equations 
assume a linear form so as to allow for the possibility 
to simply add solutions. This
linearity will be essential to the method used
 here. We refer to \cite{Bentivegna.EtAl:2018} 
for a recent comprehensive review of the
expectations and achievements connected with
lattice cosmology. More specifically, we refer 
to \cite{Clifton.EtAl:2012} for an instructive
application to the backreaction problem, to
\cite{Liu:2015} for an extensive study of redshifts and inregrated Sachs-Wolfe effects,
and \cite{Bentivegna.EtAl:2017} for a general 
discussion of light-propagation in lattice cosmology.

In \cite{LindquistWheeler} and its follow-up papers, the requirement of regularity of the lattice formed by the sites of the 
black-holes was explicitly imposed. A first relaxation from strict regularity was considered in \cite{Cliffton.Durk:2017} in relation to 
structure formation and back-reaction. Their generalisation still started from one of 
the six regular lattices, but then allowed 
to ``explode'' each black hole into a cluster of
other black holes in a special way that 
maintains overall statistical homogeneity 
and isotropy. Our method presented in this paper
can be seen as a significant generalisation of
theirs, resting on a novel application of Lie
sphere-geometry, that so far does not seem to 
have enjoyed any application to cosmological model-building whatsoever. The method itself,
the foundations of which we shall explain in 
the next section, is certainly very powerful,
though the extent to which it may profitably
applied in cosmology remains to be seen. 
As an illustrative example, we include a comparison between special black-hole
configurations that we called ``unifoamy'' 
in lattice- and swiss-cheese cosmology.
This paper is based in parts on \cite{Fennen2017}.

\section{Lie Sphere-Geometry and Apollonian Packings}
In this section we wish to acquaint the 
reader with  the geometric ideas behind
\emph{Lie sphere-geometry} and its power
to study and construct configurations 
of (round) spheres isometrically embedded in Riemannian manifolds of constant-curvature. 
As the name suggests, the geometric ideas were 
first introduced by Sophus Lie (1842-1899), in fact in 
his doctoral thesis \cite{LiePHD}. 
Our presentation will follow modern terminology and notation. As already stressed, this 
geometric method has -- quite surprisingly and
to the best of our knowledge -- not been 
employed in the general-relativistic 
initial-value problem and hardly ever in astrophysics and cosmology. The only two 
notable exceptions we are aware of concern 
the statistics of craters on planetary bodies 
\cite{GibbonsEtAl_LieSG:2013} and the 
statistics of cosmological voids \cite{GibbonsEtAl_LieSG:2014}.%
\footnote{We thank Marcus Werner for pointing 
out these references.} 
In our paper we will use it to systematically 
construct initial data for Einstein's 
field equations applied to lattice cosmology.
  
Let us now explain in some more detail those 
aspects of Lie sphere-geometry that are of 
interest to us and which we reformulate and 
amend according to our needs.  A standard 
mathematical textbook on Lie sphere-geometry 
is by T.\,E.\,Cecil \cite{cecil1992lie}, which 
contains much -- but not all -- of what we 
say in its first chapters. The central object 
in Lie sphere-geometry is the configuration space
of spheres which,  as we will see discuss in 
detail, turns out to be an old friend of all 
relativists.

\subsection{DeSitter space as configuration 
space for spherical caps, or oriented hyperspheres, within spheres}
\begin{figure}[h]
\centering
\includegraphics[width=0.5\columnwidth]{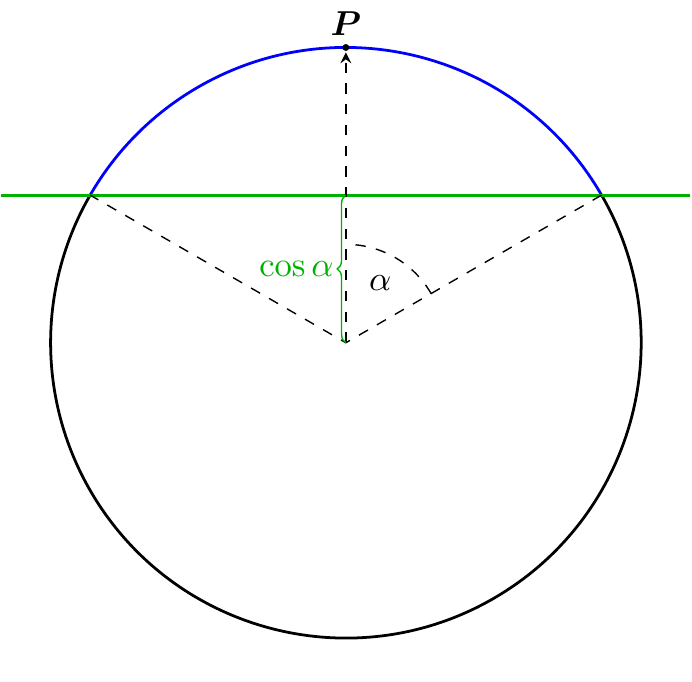}
\caption{Spherical cap in $n=1$ dimensions with centre 
$\bi{P}$ and radius $ \alpha $. The $n=2$ case is 
obtained by rotating the figure about the vertical 
symmetry axis.}
\label{fig:SphericalCap}
\end{figure}

Throughout we often consider the real vector 
space $\mathbb{R}^{n+1}$ together with its
Euclidean canonical inner product. Elements 
in $\mathbb{R}^{n+1}$ are denoted by bold-faced 
letters, like  $\bi{X}$ and  $\bi{P}$, and their inner product $\bi{X} \cdot \bi{P}$ is 
defined as usual. The inner product defines a 
norm 
$\Vert\bi{X}\Vert:=\sqrt{\bi{X}\cdot\bi{X}}$. 
The $n$-sphere of unit-norm vectors in $\mathbb{R}^{n+1}$ is  
\begin{equation}
\label{eq:DefUnitSphere}
S^n=\left\{\bi{X}\in\mathbb{R}^{n+1}: 
\left\|\bi{X}\right\|=1\right\}\,.
\end{equation}
The geodesic distance $\Lambda\bigl(\bi{X},\bi{P}\bigr)\in[0,\pi]$ 
between the two points $\bi{X}$ and $\bi{P}$ on $S^n$
is given by 
\begin{equation}
\label{eq:DefGeodDistance}
\Lambda\bigl(\bi{X},\bi{P}\bigr):=\arccos\bigl(\bi{X}\cdot\bi{P}\bigr)\,.
\end{equation}
A \emph{spherical $\alpha$-cap} on $S^n$, with $\alpha\in(0,\pi)$, 
centered at $\bi{P}\subset S^n$ is the set of all points 
$\bi{X}\in S^n $ whose geodesic distance from $\bi{P}$ 
is less or equal to $\alpha$. Hence these points 
satisfy 
\begin{equation} 
\label{eqn:IntersectingPlane}
\bi{X}\cdot\bi{P}\geq\cos\alpha\,.
\end{equation}
It should be read as an equation describing the 
intersection between the  half-space 
$\{\bi{X}\in\mathbb{R}^{n+1}:\bi{X}\cdot\bi{P}\geq\cos\alpha\}$
with $S^n$. See \fref{fig:SphericalCap} for an illustration 
of the cases $n=1,2$. 

Now, the central idea of Lie sphere-geometry is 
to regard $S^n$ not as subset of $\mathbb{R}^{n+1}$ 
endowed with the Euclidean inner product, but rather 
as subset of $(n+2)$-dimensional Minkowski space 
$\mathbb{R}^{1,n+1}$, i.e. the vector space 
$\mathbb{R}^{n+2}$ endowed with a non-degenerate symmetric 
bilinear form of signature $(1,n+1)$, the so-called Minkowski
metric or Minkowski inner product, which in the 
``mostly-plus-convention'' that we shall use here is given 
by
\begin{equation}
\label{eq:DefMinkInnerProduct}
\langle\bxi_1,\bxi_2\rangle:=
-\xi_1^0\xi_2^0+\sum_{a=1}^{n+1}\xi_1^a\xi_2^a\,.	
\end{equation}
Hence spacelike vectors have positive and timelike 
vectors have negative Minkowski square.

The embedding of $S^n\subset\mathbb{R}^{n+1}$ into  
$\mathbb{R}^{1,n+1}$ is then given by regarding 
$\mathbb{R}^{n+1}$ as affine spacelike hyperplane of 
constant time (first coordinate in $\mathbb{R}^{1,n+1}$) 
equal to 1. Then
\begin{equation}
\label{eq:EmbeddingSphereMinkSpace}
\mathbb{R}^{n+1}\supset S^n\ni\bi{X}\longmapsto 
\bxi:=\left(1,\bi{X}\right) \in \mathbb{R}^{1,n+1}\,.
\end{equation}
Obviously $\langle\bxi,\bxi\rangle=0$, so that 
$S^n\subset\mathbb{R}^{1,n+1}$ is the intersection of the 
constant-time hyperplane with the future light-cone with 
vertex at the origin. This intersection is also called 
the \emph{M\"obius sphere}. 

Like above, a spherical cap on the M\"obius sphere can be 
obtained by intersecting the latter with a half space. 
But now the half space is such that its boundary hyperplane, 
which is timelike,  contains the origin of $\mathbb{R}^{1,n+1}$. 
Hence we can rewrite \Eref{eqn:IntersectingPlane} as
\begin{equation}
\label{eqn:CapEquation}
\langle \bxi, \bomega \rangle \geq 0\,,
\end{equation}
where (recall $\csc(x)=1/\sin(x)$)
\begin{equation}
\label{eqn:LieCap}
\bomega=\bigl(\cot(\alpha), 
\bi{P}\csc(\alpha)\bigr)
\end{equation}
is a normalized spacelike vector, i.e. 
$\langle\bomega,\bomega\rangle=1$, which is 
Minkowski-perpendicular to the boundary hyperplane of the 
half-space and oriented such that it points into the 
interior of the half-space. It is sometimes referred to as 
\emph{Lie (sphere) vector}. It establishes a 
bijection between the set of spherical caps of non-zero 
radius in $S^n$ -- equivalently the set of oriented 
$(n-1)$ spheres (hyperspheres) of non-zero radius in 
$S^n$ -- and the set of unit spacelike vectors in 
$\mathbb{R}^{1,n+1}$. The later is just the one-sheeted
 timelike unit hyperboloid in $(n+2)$-dimensional 
Minkowski space, known to relativists as 
$(n+1)$-dimensional deSitter space of unit 
radius, which we denote by $\mathrm{dS}^{n+1}$. 
It thus assumes the role of the configuration 
space of spherical caps - or oriented hyperspheres - 
in $S^n$. Remarkably, this configuration space is
itself endowed with a natural \emph{Lorentzian}
geometry that it inherits from being imbedded into 
Minkowski space and that is well known to relativists. 
Indeed, if we restrict the $(1,n+1)$ Minkowski metric
\begin{equation}
\label{eq:MinkMetric}
{\bfeta}=
-\mathbf{d}\omega^0\otimes\mathbf{d}\omega^0
+\sum_{a=1}^{n+1}\mathbf{d}\omega^a\otimes\mathbf{d}\omega^a
\end{equation}
to the tangent bundle of the embedded timelike 
hyperboloid in the parametrisation \eref{eqn:LieCap}, 
where $\bi{P}$ is normalised, so that 
$\bi{P}\cdot\mathbf{d}\bi{P}=0$, we immediately get
\begin{equation}
\label{eq:DeSitterMetric-1}
\bi{g}^{\mathrm{dS}^{(n+1)}}=\csc^2(\alpha)
\biggl(
-\mathbf{d}\alpha\otimes\mathbf{d}\alpha
+\bi{g}^{S^n}\biggr)\,.
\end{equation}
Here $\bi{g}^{S^n}$ denotes the standard round 
metric of the unit $n$-sphere $S^n$ given by 
restricting $\sum_{i=1}^{n+1}\mathbf{d}\bi{P}_i\otimes\mathbf{d}\bi{P}_i=:
\mathbf{d}\bi{P}\dot\otimes\mathbf{d}\bi{P}$
to the $n$-sphere $\Vert\bi{P}\Vert=1$. 
Replacing $\alpha\in(0,\pi)$ by $t\in(-\infty,\infty)$ 
according to the reparametrisation 
\begin{equation}
\label{eq:DeSitterRepara}
t=t(\alpha):=
\left\{
\begin{array}{ll}
-\,\mathrm{arccosh}\bigl(\csc(\alpha)\bigr)
&\mathrm{for}\quad 0<\alpha\leq\pi/2\\
 +\, \mathrm{arccosh}\bigl(\csc(\alpha)\bigr)
&\mathrm{for}\quad \pi/2\leq\alpha<\pi
\end{array}\right.
\end{equation}
leads to the well known form of the deSitter metric 
used for $n=3$ in standard relativistic cosmology:
\begin{equation}
\label{eq:DeSitterMetric-2}
\bi{g}^{\mathrm{dS}^{(n+1)}}=
-\mathbf{d}t\otimes\mathbf{d}t
+\cosh^2(t)\,\bi{g}^{S^n}\,.
\end{equation}
Note that the function on the right-hand side of \eref{eq:DeSitterRepara} maps the interval $(0,\pi)$ 
strictly increasing and differentiably onto 
$(-\infty,\infty)$. Indeed, the derivative of 
$t(\alpha)$ is just $t'(\alpha)=\csc(\alpha)$ for all 
$0<\alpha<\pi$. 

\begin{figure}[tb]
\centering
\includegraphics[trim={2cm 2cm 1.5cm 1cm}, 
 clip, width=0.6\columnwidth]{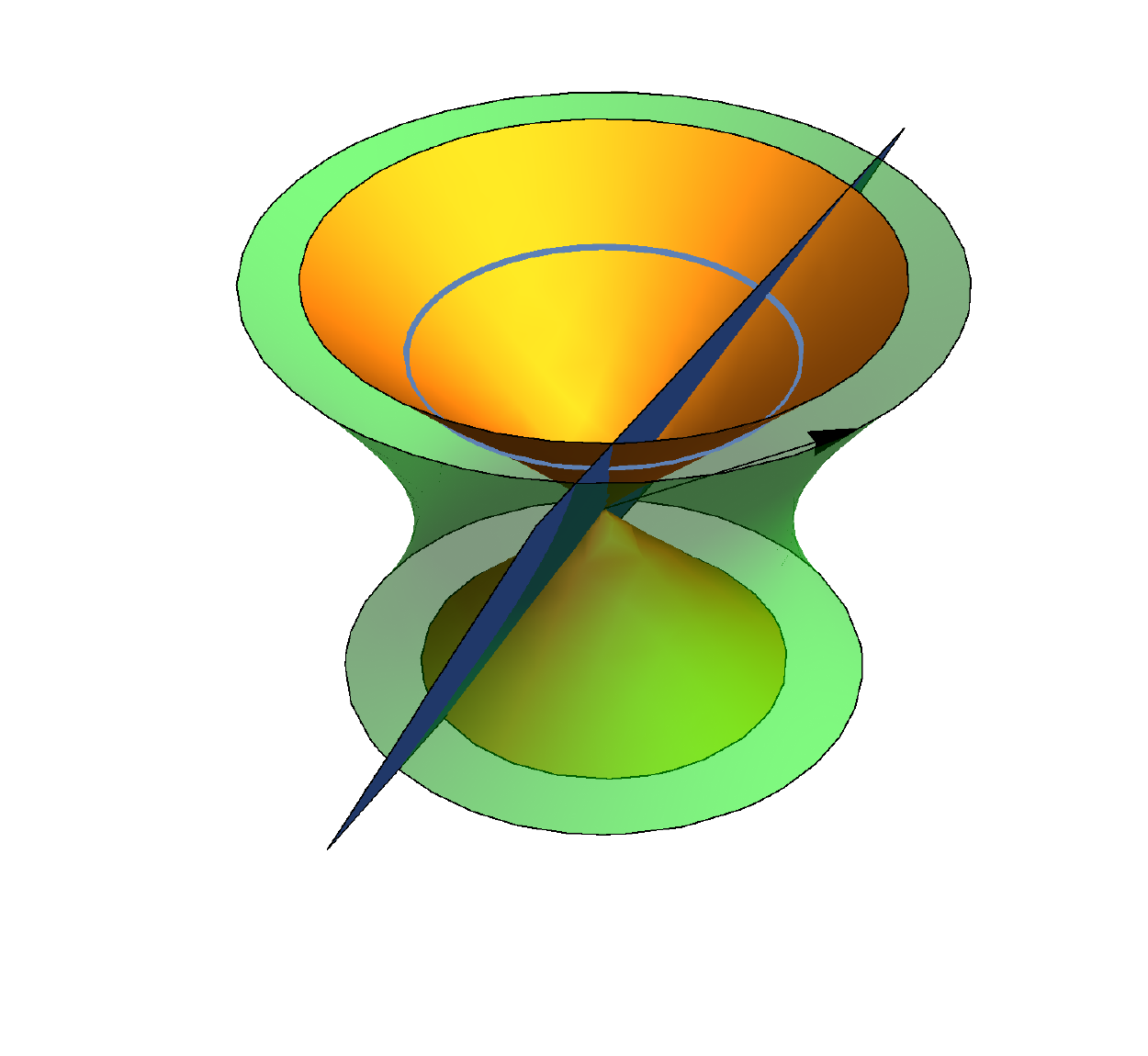}
\caption{\label{fig:LieSphere}%
Illustration of the bijection between spherical caps
or oriented hyperspheres in the M\"obius sphere 
$S^n$ and points on $\mathrm{dS^{(n+1)}}$, here for 
$n=1$. The picture shows various geometric objects 
embedded into $(1+2)$-dimensional Minkowski space: 
The $2$-dimensional light-cone is depicted in yellow, 
the $2$-dimensional hyperboloid of unit spacelike 
vectors, i.e.  $2$-dimensional deSitter space, 
in green. The Moebius sphere is the intersection 
of the light-cone with an affine hyperplane (not 
shown in the diagram) of constant unit time, here
depicted by the light-blue circle. Finally, the 
oriented timelike hyperplane through the 
origin is in dark-blue and its (oriented) normal by 
the black arrow, denoted by $\bomega$ in the text, 
whose tip defines a point on the green hyperboloid. 
This point uniquely defines a spherical cap of -- or 
oriented hypersphere in -- the Moebius sphere. Note 
that the closure of the complement of the spherical 
cap is also a spherical cap bounded by the same but 
oppositely oriented hypersphere, and both are 
represented by $-\bomega$.}
\end{figure}

In this fashion the set of spherical caps in $S^n$ 
is not only put into bijective correspondence with 
points in $\mathrm{dS}^{n+1}$, but is also endowed
with the structure of a maximally symmetric 
Lorentzian manifold with metric 
$\bi{g}^{\mathrm{dS}^{(n+1)}}$, the geometry of which 
turns out to be very useful indeed, with many and 
sometimes  surprising applications. For example, in 
$n=2$ and $n=3$
dimensions, the volume form induced by this metric has 
been used for statistical discussions of distributions 
of planetary craters in \cite{GibbonsEtAl_LieSG:2013} 
and cosmic voids in \cite{GibbonsEtAl_LieSG:2014}, respectively. In \fref{fig:LieSphere} we illustrate 
once more the geometric objects underlying this 
bijective correspondence between spherical caps 
of -- or oriented hyperspheres in --  the M\"obius 
sphere $S^n$ and deSitter space $dS^{(n+1)}$ in the 
case $n=1$.

As regards the Lorentzian signature of 
$\bi{g}^{\mathrm{dS}^{(n+1)}}$, note that changing 
the location of the spherical cap's centre while 
keeping the radius fixed corresponds to a spacelike 
motion in configuration space, while a change in 
radius with fixed centre corresponds to a timelike 
motion. Increasing cap radii correspond to increasing 
$\alpha$ and hence increasing $t$ according to \eref{eq:DeSitterRepara}. The set of caps carries a 
natural partial-order relation given by inclusion. 
It is geometrically obvious that a cap with centre 
$\bi{P}$ and geodesic radius $\alpha$ is properly 
included in another one parametrised by $\bi{P}'$ 
and $\alpha'$, if and only if the geodesic distance 
between the centres is less than, or equal to, the 
difference $\alpha'-\alpha$ of their geodesic radii. 
As the geodesic distance between  $\bi{P}$ and 
$\bi{P}'$ is measured by $\bi{g}^{S^n}$ in \eref{eq:DeSitterMetric-1}, the latter condition
of proper containment is seen to be equivalent to 
the condition that the corresponding points $\bomega$
and $\bomega'$ on $\mathrm{dS^{(n+1)}}$ are timelike
or lightlike separated with $\bomega'$ to the future
of $\bomega$ in the time orientation given by 
increasing $t$. This shows that the set-theoretic 
partial-order relation of spherical caps given by 
containment just corresponds to the partial-order
relation on $\bigl(\mathrm{dS^{(n+1)}},\bi{g}^{\mathrm{dS}^{(n+1)}}\bigr)$ given by causality. More precisely, 
$\bomega'$ lies to the causal future of  $\bomega$ 
if the cap corresponding to $\bomega$ is properly 
contained in the cap corresponding to $\bomega'$. 
This causal separation is timelike if the smaller
cap is properly contained in the interior of the larger
one, and lightlike if the boundary spheres of the caps 
just touch at one point. (We will come back to this 
order relation in more detail when we discuss the 
images of caps of $S^n$ under stereographic projection 
in $\mathbb{R}^n$, where they become balls.)  
It is intriguing that, in this way, Lie sphere-geometry 
provides a natural link between causal- and cap- or 
``sphere-orders''. In fact, this relation is inherent 
in the discussion of sphere orders in 
\cite{Brightwell.Winkler:1989}%
\footnote{We thank Fay Dowker for pointing out this
reference.}, 
the motivation of which came from causal orders,
however without relating it to Lie sphere-geometry. 

\subsection{Balls and oriented hyperspheres in flat euclidean space}
\label{sec:BallsHyperspheresEuclidean}
The foregoing construction also applies to balls, or 
oriented hyperspheres, in flat euclidean space 
$\mathbb{R}^n$ if suitably generalised. 
To see this we regard $S^n$ as one-point 
compactification of $\mathbb{R}^n$. The point
added to  $\mathbb{R}^n$ is called ``infinity'' 
and denoted by $\infty$. The set 
$\mathbb{R}^n\cup\{\infty\}$ is topologised 
in such a way that complements of compact sets in 
$\mathbb{R}^n$ become open neighbourhoods of 
$\infty$, which makes $\mathbb{R}^n\cup\{\infty\}$
homeomorphic to $S^n$. A homeomorphism is given 
by inverse stereographic projection centred at, say, 
the ``south pole'' $(0,\dots,0,-1)$; compare \eref{eq:Appendix2-2}:
\begin{equation}
\label{eqn:StereographicProjection}
\mathbb{R}^n\ni\bi{x}\,\mapsto\,\bi{X} 
= \left(\frac{2 \bi{x}}{1 + \bi{x}^2}, 
\frac{1-\bi{x}^2}{1 + \bi{x}^2}\right) 
\in S^n\subset\mathbb{R}^{n+1}\,.
\end{equation}
An important property of stereographic projections is 
that balls in $\mathbb{R}^n$ are mapped to spherical 
caps in $S^n$.\footnote{Images of balls in 
$\mathbb{R}^n$ under \eref{eqn:StereographicProjection}
are spherical caps not containing $\infty$. Spherical
caps containing $\infty$ in their interior or on their 
boundary are images under 
\eref{eqn:StereographicProjection} of closures 
of complements of balls and images of half-spaces,
 respectively. This will be further discussed below.}
 Consequently we can use Lie
sphere-geometry to also describe the configurations of 
balls, or oriented hyperspheres, in $\mathbb{R}^n$. 
As before, the $n$-sphere can be embedded into 
$\mathbb{R}^{(1,n+1)}$ (to become the M\"obius
sphere) via $\bxi=(1,\bi{X})$, where 
$\bi{X}$ is a unit vector in euclidean 
$\mathbb{R}^{n+1}$, which is now to be expressed 
through $\bi{x}$ according to \eref{eqn:StereographicProjection}. 
A ball in $\mathbb{R}^n$ with centre $ \bi{p} $ and 
radius $r>0$ is defined as the set of all points 
$\bi{x}$ satisfying
\begin{equation}
\label{eq:BallInRn-1}
\left( \bi{x} - \bi{p} \right)^2 \leq r^2.
\end{equation}
A short calculation shows that this is equivalent 
to
\begin{equation}
\label{eq:BallInRn-2}
\langle \bxi, \bomega \rangle \geq 0,
\end{equation}
where
\begin{equation} 
\label{eq:BallCoordinates}
\bomega = \left( \frac{1 + \bi{p}^2 - r^2}{2 r}, 
\frac{\bi{p}}{r}, \frac{1 - \bi{p}^2 + r^2}{2 r} \right)\in\mathbb{R}^{(1,n+1)}
\end{equation}
is a spacelike unit vector in $(n+2)$-dimensional 
Minkowski space $\mathbb{R}^{(1,n+1)}$. 

Note that the closure of the complement of the ball 
described 
by \eref{eq:BallInRn-1} is described by the reversed 
inequality, $\left(\bi{x}-\bi{p}\right)^2\geq r^2$, 
hence by $\langle\bxi,\bomega\rangle\leq 0$ instead 
of \eref{eq:BallInRn-2}. Consequently, the complement 
of a ball represented by $\bomega$ is represented 
by $-\bomega$, just as before. We can use the same 
representation \eref{eq:BallCoordinates} if we associate a 
negative radius $ r<0 $ to these sets. Hence, a Lie vector 
$\bomega$ can represent either a ball (with positive and 
negative radius) using \eref{eq:BallCoordinates} or a 
spherical cap via \eref{eqn:LieCap}. 

\begin{figure}[h]
	\centering
	\includegraphics[trim={0 3cm 0 1cm}, clip, width=0.7\columnwidth]{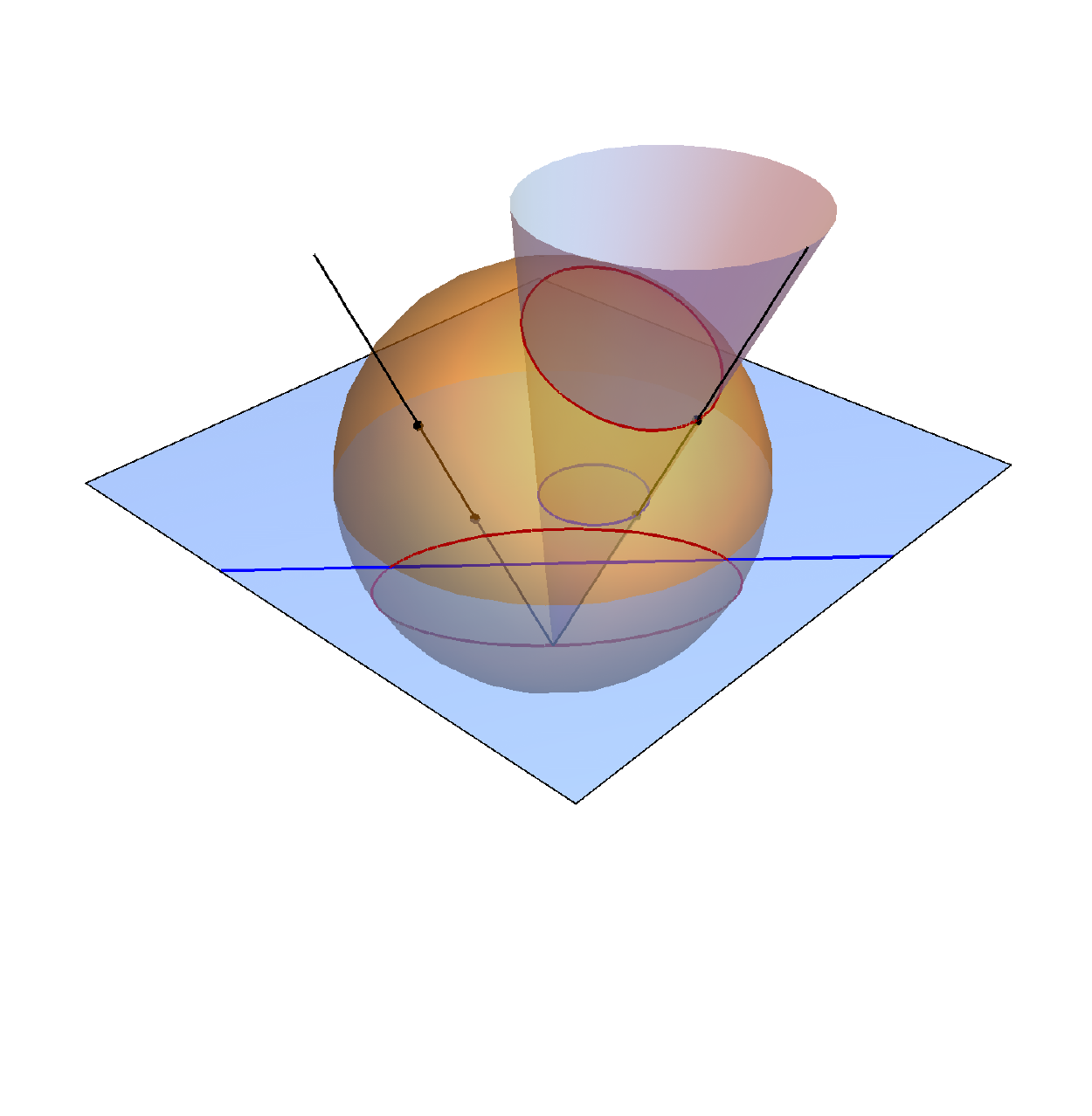}
	\caption{A Lie vector describes either a spherical cap or a ball/half-space.}
	\label{fig:Projection}
\end{figure}

However, not all points on de\,Sitter space can be
 parametrised by \eref{eq:BallCoordinates}; we are 
missing those which are parametrised by
\begin{equation}
\label{PlaneCoordinates}
\bomega = \left( -d, \bi{n}, d \right)\,,
\end{equation}
where $\bi{n}^2=1$. If we consider the scalar product 
$\langle\bxi,\bomega\rangle\geq 0$, we obtain
\begin{equation}
	\bi{n} \cdot \bi{x} \geq d\,.
\end{equation}
This is a half-space in $ \mathbb{R}^n $ with a 
boundary plane with outward-pointing normal $ \bi{n} $ 
and distance $d$ from the origin. It can be shown 
that these half-spaces correspond to caps containing 
the south pole on their boundary. Hence, half-spaces 
can be interpreted as balls just touching infinity
with their boundary. Altogether, there is a bijective
 correspondence between spherical caps on $S^n$ on 
one side, and balls, their complements, and 
half-spaces in $\mathbb{R}^n$ on the other. We will 
use this fact to visualise caps on the 3-sphere as 
the corresponding objects in $\mathbb{R}^3$. 
The two-dimensional case is shown in \fref{fig:Projection}.

\subsection{Intersecting caps,  or oriented 
hyperspheres, and Descartes configurations}

From \eref{eq:BallCoordinates} we can easily calculate 
the Minkowskian inner product between two vectors 
$\bomega_1$ and $\bomega_2$ in $\mathbb{R}^{(1,n+1)}$
representing balls with parameters $(\bi{p}_1,r_1)$ and 
$(\bi{p}_2,r_2)$, respectively. The result is 
\begin{equation}
	\label{eq:BallCoordinates-InnerProduct}
\langle\bomega_1, \bomega_2\rangle=
\frac{r_1^2+r_2^2-\Vert\bi{p_1}-\bi{p_2}\Vert^2}{2r_1r_2} =
\pm 1+\frac{(r_1\mp r_2)^2-\Vert\bi{p}_1-\bi{p}_2\Vert^2}{2r_1r_2}\,.
\end{equation}
Here the second equality holds either with both 
upper or both lower signs in the terms on the 
right-hand side. It immediately shows that 
$\langle\bomega_1, \bomega_2\rangle\in[-1,1]$ iff 
\begin{equation}
	\label{eq:IntersectionCondition}
\vert r_1-r_2\vert\leq\Vert\bi{p}_1-\bi{p}_2\Vert\leq\vert r_1+r_2\vert\,,
\end{equation}
with $\langle\bomega_1, \bomega_2\rangle=-1$
for $\Vert\bi{p}_1-\bi{p}_2\Vert=\vert r_1+r_2\vert$ and 
$\langle\bomega_1, \bomega_2\rangle=1$
for $\Vert\bi{p}_1-\bi{p}_2\Vert=\vert r_1-r_2\vert$.
It is geometrically clear that if 
$\Vert\bi{p}_1-\bi{p}_2\Vert>\vert r_1+r_2\vert$, i.e.
if $\langle\bomega_1, \bomega_2\rangle<-1$,
the balls represented by $\bomega_1$ and $\bomega_2$
are disjoint; and that they just touch at a single 
boundary point, with oppositely pointing normals, if
$\Vert\bi{p}_1-\bi{p}_2\Vert=\vert r_1+r_2\vert$, i.e.
if $\langle\bomega_1, \bomega_2\rangle=-1$. Moreover, if 
$\Vert\bi{p}_1-\bi{p}_2\Vert<\vert r_1-r_2\vert$, i.e. 
if $\langle\bomega_1, \bomega_2\rangle>1$, then either 
the ball represented by $\bomega_1$ is entirely contained
in the interior of that represented by $\omega_2$
(case  $r_1<r_2$) or vice versa (case $r_2<r_1$). For  
$\Vert\bi{p}_1-\bi{p}_2\Vert=\vert r_1-r_2\vert$, i.e.
$\langle\bomega_1, \bomega_2\rangle=1$, one ball is 
contained in the other with their boundaries touching
at a single point with parallely pointing normals. 

This shows that \eref{eq:IntersectionCondition} is 
just the necessary and sufficient condition for the 
oriented boundary spheres of the balls to intersect. 
The angle between the normals at an intersection 
point is clearly independent of the intersection point and 
referred to as the \emph{intersection angle} of the 
spheres. Applying the law of cosines to the triangle 
with vertices $\bi{p}_1$, $\bi{p}_2$, and an 
intersection point of the spheres with radii $r_1$ and 
$r_2$ centered at $\bi{p}_1$ and $\bi{p}_2$, 
respectively, immediately gives 
\begin{equation}
\label{eq:IntersectionAngle-1}
\Vert\bi{p}_1-\bi{p}_2\Vert^2
=r_1^2+r_2^2-2r_1r_2\cos(\gamma_{12})\,,
\end{equation}
where $\gamma_{12}$ is the angle of the triangle at the 
intersection vertex, which is just the intersection angle
of the spheres. Using \eref{eq:IntersectionAngle-1} in
the first equality of \eref{eq:BallCoordinates-InnerProduct}
leads to the simple formula 
\begin{equation}
\label{eq:IntersectionAngle-2}
\langle\bomega_1, \bomega_2\rangle=\cos(\gamma_{12})\,.
\end{equation}
In particular, $\langle\bomega_1, \bomega_2\rangle=0$
means that the spheres intersect orthogonally, whereas 
$\langle\bomega_1, \bomega_2\rangle=1$ and $\langle\bomega_1, \bomega_2\rangle=-1$ means that the spheres just touch 
tangentially with one containing the other in the first,  
and disjoint interiors in the second case.

On the 
$n$-sphere it is possible to find sets of (at most) 
$ n+2 $ pairwise tangent caps. Such a set 
$\{ \bomega_a: a=1,\dots,n+2\} $ is called a
\emph{Descartes set}, in view of Descartes' circle
theorem for four circles in flat two-dimensional 
space $\mathbb{R}^2$, giving a relation between 
the radii. The generalisation to higher dimensions 
was given by Frederick\,Soddy \cite{Soddy1936} and Thorold \,Gosset \cite{Gosset1937} in form of poems! There are several
formulae which also include the centres and extensions 
to other constant-curvature spaces \cite{Lagarias2002}.
Lie sphere-geometry provides an elegant and powerful 
unification of all these results.

Indeed, the caps of a Descartes set have to satisfy
\begin{equation}
\label{eq:DescartesCondition}
\langle \bomega_a, \bomega_b \rangle 
=2\delta_{ab} - 1
\end{equation}
because $\langle\bomega_a,\bomega_a\rangle=1$
for all Lie vectors and 
$\langle\bomega_a,\bomega_b\rangle=-1$ if $a\neq b$ 
as condition for touching at one point. 
Writing the Descartes set as a square 
$(n+2)\times(n+2)$ matrix $\bi{W}$ whose 
rows are the components of the 
vectors $\bomega_{a}$, that is, 
$\bi{W}^\top=\left( \bomega_{1},\dots,\bomega_{n+2}\right)$, 
we obtain the equivalent to \eref{eq:DescartesCondition}:
\begin{equation} 
\label{eqn:InverseDescartesTheorem}
\bi{W} \bfeta \bi{W}^T = \bi{G}\,,
\end{equation}
where $\bfeta=\mathrm{diag}(-1,1,\dots,1)$ is 
the Minkowski metric and 
$\bi{G}_{ab}=2 \delta_{ab} - 1 $. 
Simply inverting \eref{eqn:InverseDescartesTheorem} 
leads to (matrices  with components
$(2\delta_{ab}-1)$ are non-singular in dimensions higher than two):

\begin{equation} 
\label{eqn:DescartesTheorem}
\bi{W}^T \bi{G}^{-1} \bi{W} = \bfeta\,,
\end{equation}
which is known as the 
\emph{unified generalised Descartes theorem} 
containing formulae for centres as well as 
radii~\cite{Lagarias2002}. We shall be no more explicit at this point. But we think that the simple half-page argument leading to \eref{eqn:DescartesTheorem}, comprising the most general statement on the 
general Descartes' theorem, impressively demonstrates the ability of Lie 
sphere-geometry.

Using the inverse of $\bi{G}$, we can 
define a set of \emph{dual caps} (compare \cite{Soderberg1992}) $\btau_a$ via
\begin{equation}
\label{eqn:DualCap}
\btau_a:=\kappa\,\sum_{b=1}^{n+2}(\bi{G}^{-1})_{ab}\,\bomega_b\,,
\end{equation}
where $\kappa^2=\frac{2n}{n-1}$
is needed for normalisation 
such that $ \langle \btau_a, \btau_a \rangle = 1 $. The 
components of the inverse matrix $ \bi{G}^{-1} $ 
are given by 
$(\bi{G}^{-1})_{ab}=\frac{1}{2}\left(\delta_{ab}-\frac{1}{n}\right)$. 
The dual caps satisfy
\begin{eqnarray}
\langle\btau_a, \btau_b\rangle
=\frac{n\,\delta_{ab} - 1}{n - 1} \leq 1\,, \\
\langle\btau_a, \bomega_b\rangle=\kappa\,\delta_{ab}\,,
\end{eqnarray}
showing that the cap $\btau_a$ is orthogonal to 
all caps $\bomega_b$, $b\neq a$. Furthermore, 
the dual caps overlap in more than two dimensions, 
as the first equation shows.

\subsection{Apollonian groups and the generation of apollonian packings}
\begin{figure}[htb]
	\centering
	\includegraphics[width=0.6\columnwidth]{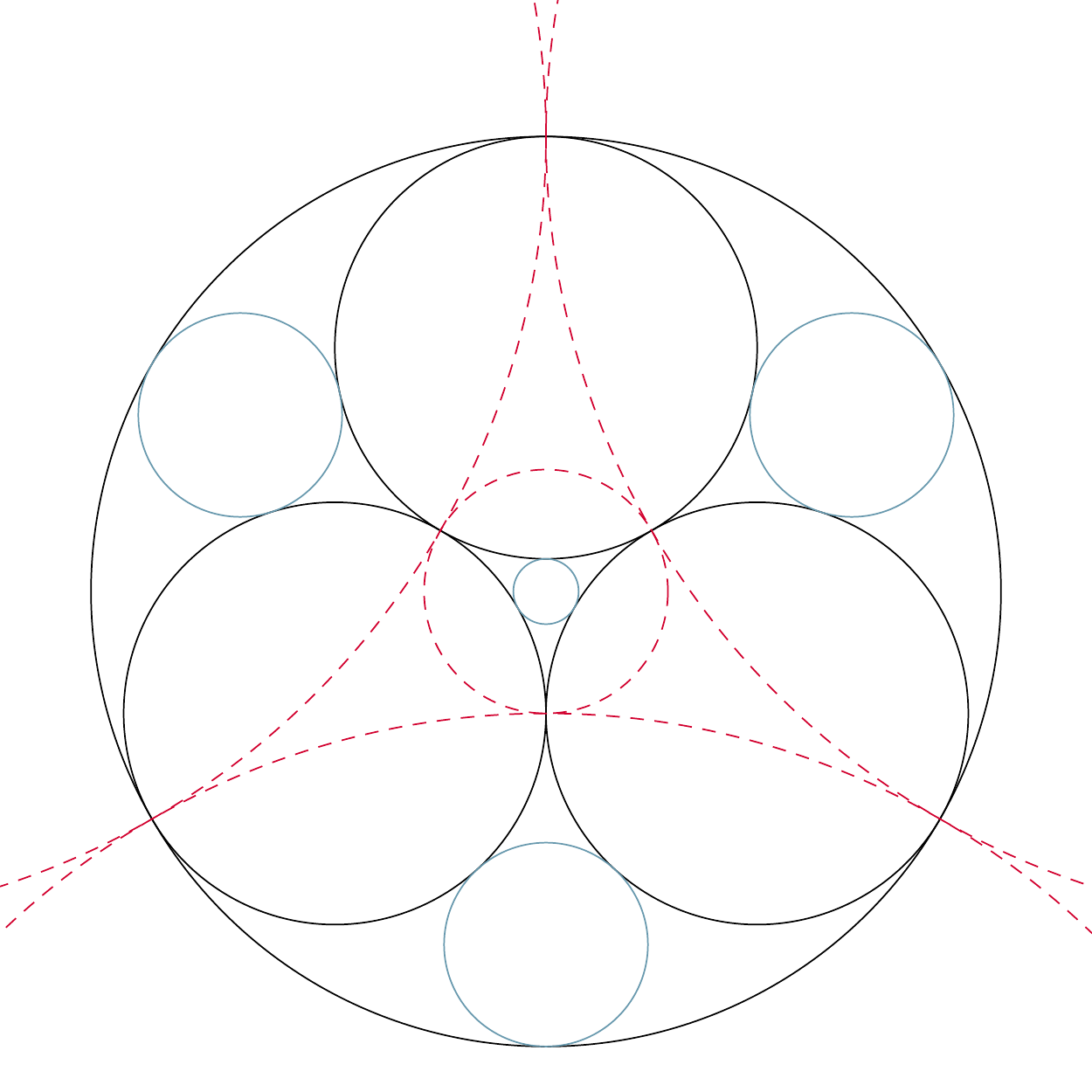}
	\caption{First iteration of a two-dimensional Apollonian packing: initial set in black, dual set in red and reflected set in blue.}
	\label{fig:FirstIteration}
\end{figure}

The dual set just introduced can now be used to 
construct new spheres tangent to a given 
Descartes set. For this we define the mapping 
$\bi{I}_{\bi{\btau}_a} $ acting on the set of 
all Descartes sets via
\begin{equation} 
\label{eq:HyperplaneReflection}
\bomega_b' 
= \bi{I}_{\bi{\btau}_a}\bomega_b=\bomega_b-2
  \left\langle \bomega_b, \btau_a \right\rangle 
\btau_a\,.
\end{equation}
In Minkowski space $\mathbb{R}^{(1,n+1)}$ it 
corresponds to a reflection in the timelike 
hyperplane with unit normal $\btau_a$. Hence 
we have 
$\bomega_b'=\bi{I}_{\bi{\btau}_a}\bomega_b=\bomega_b$ 
if $b\neq a$, since $\langle \bomega_b,\btau_a\rangle=0$,
since $\bomega_b$ lies in the hyperplane of reflection, 
which is clearly pointwise fixed. It can be easily 
verified that the set 
$\left\{ \bi{I}_{\bi{\btau}_a} \bomega_a, \bomega_b : b \neq a \right\}$ forms a new Descartes set. Being reflections, 
the maps $\bi{I}_{\bi{\btau}_a} $ clearly preserve the 
Minkowski inner product,  i.e. they are Lorentz 
transformations, so that $\langle \bi{I}_{\bi{\btau}_a}
\bomega_b, \bi{I}_{\bi{\btau}_a} \bomega_c \rangle =
\langle \bomega_b, \bomega_c \rangle $. It can be shown 
that these maps also act on $S^n$ and $\mathbb{R}^n$ by 
considering their points as spheres of radius zero. 
The hyperplane reflection $\bi{I}_{\bi{\btau}_a}$ then 
becomes an inversion on the sphere that is the boundary 
of the ball represented by $\btau_a$. Let us recall that in $\mathbb{R}^n$ the map 
that inverts at a sphere with centre 
$\bi{p}$ and radius $r$ is simply given by 
\begin{equation}
\label{eq:SphereInversion}
\bi{x}\mapsto \bi{x}' := \bi{p} 
+\frac{r^2}{\Vert\bi{x} 
- \bi{p} \Vert^2} 
\left( \bi{x} - \bi{p} \right)\,.
\end{equation}
In passing we make the cautionary remark 
that whereas inversions map balls and spheres to balls and spheres, their
centres will not be images of each other.
For us a truly remarkable property 
will be important: namely that this correspondence of maps relates the 
non-linear inversion 
\eref{eq:SphereInversion} to the \emph{linear} 
hyperplane reflection \eref{eq:HyperplaneReflection}.
This will simplify calculations considerably and once 
more exemplifies the power of Lie sphere-geometry, 
which gives a unified description for the flat and 
spherical case, which includes 
points and caps, as well as balls and 
half-spaces; see \fref{fig:FirstIteration}. 

The mapping \eref{eq:HyperplaneReflection} 
$\left\{ \bomega_a \right\} \mapsto \left\{ \bomega_a' 
= \bi{I}_{\bi{\btau}_b} \bomega_a \right\} $ 
can also be written as follows
\begin{eqnarray}
	\bomega_a' &=& \sum_c (\bi{A}_b)_{ac} \,\bomega_c = \bomega_a,  \qquad a \neq b, \label{eqn:ApollonianMatrix1} \\
	\bomega_b' &=& \sum_c (\bi{A}_b)_{ac} \,\bomega_c = - \bomega_b + \frac{2}{n-1} \sum_{c \neq b} \bomega_c, \label{eqn:ApollonianMatrix2}
\end{eqnarray}
where $ \bi{A}_b $ are the so-called \emph{Apollonian matrices}. 
For example, in two and three dimensions, $ \bi{A}_1 $ takes the form
\begin{equation}
	\bi{A}_1 = \left( \begin{array}{ccccc}
	-1 & 2 & 2 & 2 \\
	0 & 1 & 0 & 0 \\
	0 & 0 & 1 & 0 \\
	0 & 0 & 0 & 1 
	\end{array} \right), \qquad
	\bi{A}_1 = \left( \begin{array}{ccccc}
	-1 & 1 & 1 & 1 & 1 \\
	0 & 1 & 0 & 0 & 0 \\
	0 & 0 & 1 & 0 & 0 \\
	0 & 0 & 0 & 1 & 0 \\
	0 & 0 & 0 & 0 & 1 
	\end{array} \right).
\end{equation}
The group $ \mathcal{A} = \langle \bi{A}_1, \dots, \bi{A}_5 \rangle $ 
generated by the Apollonian matrices is called \emph{Apollonian group} and was 
studied in \cite{Graham1,Graham2,Graham3}. It is a sub-group 
of the automorphism group of $ \bi{G}^{-1} $, that is, 
$ \bi{A}^t \bi{G}^{-1} \bi{A} = \bi{G}^{-1} $, $ \bi{A} \in \mathcal{A} $. 
\Eref{eqn:DescartesTheorem} shows that the Apollonian group is conjugate 
to a sub-group of the Lorentz group. The inversions $ \bi{I} $ act 
from the left on $ \bi{W}^t $, whereas elements $ \bi{A} $ of 
the Apollonian group act from the left on $ \bi{W} $.

For $ n=2 $ and $ n=3 $, an orbit of the Apollonian group gives 
an ``almost-covering'' of the $ n $-sphere with non-overlapping 
spherical caps. This ceases to be true in higher dimensions 
because the Apollonian groups consists of integer matrices 
only in two and three dimensions. The residual sets of 
points not contained in any cap form fractals of Hausdorff 
dimension 1.3057 ($ n=2 $) \cite{Graham1} and 2.4739 ($ n=3 $) 
\cite{Graham3,Borkovec1994}. Since the Lorentz group acts 
transitively on the set of all Descartes sets, one might 
say that there is only one Descartes set, and consequently 
only one Apollonian packing, up to Lorentz transformations.

The advantage in using the inversions $ \bi{I}_{\bi{\btau}_b} $ 
rather than the Apollonian matrices $ \bi{A}_b $ is that the 
former can act on single caps whereas the latter can only 
act on Descartes sets $ \bi{W} $. For this reason they 
are more useful for numerical calculations of Apollonian 
packings. Note that the representation of the inversion 
matrices $ \bi{I}_{\bi{\btau}_a} $ depends on the chosen 
Descartes set, whereas the Apollonian matrices are defined 
independently of any such choice.

\begin{figure}
	\centering
	\includegraphics[width=0.7\columnwidth]{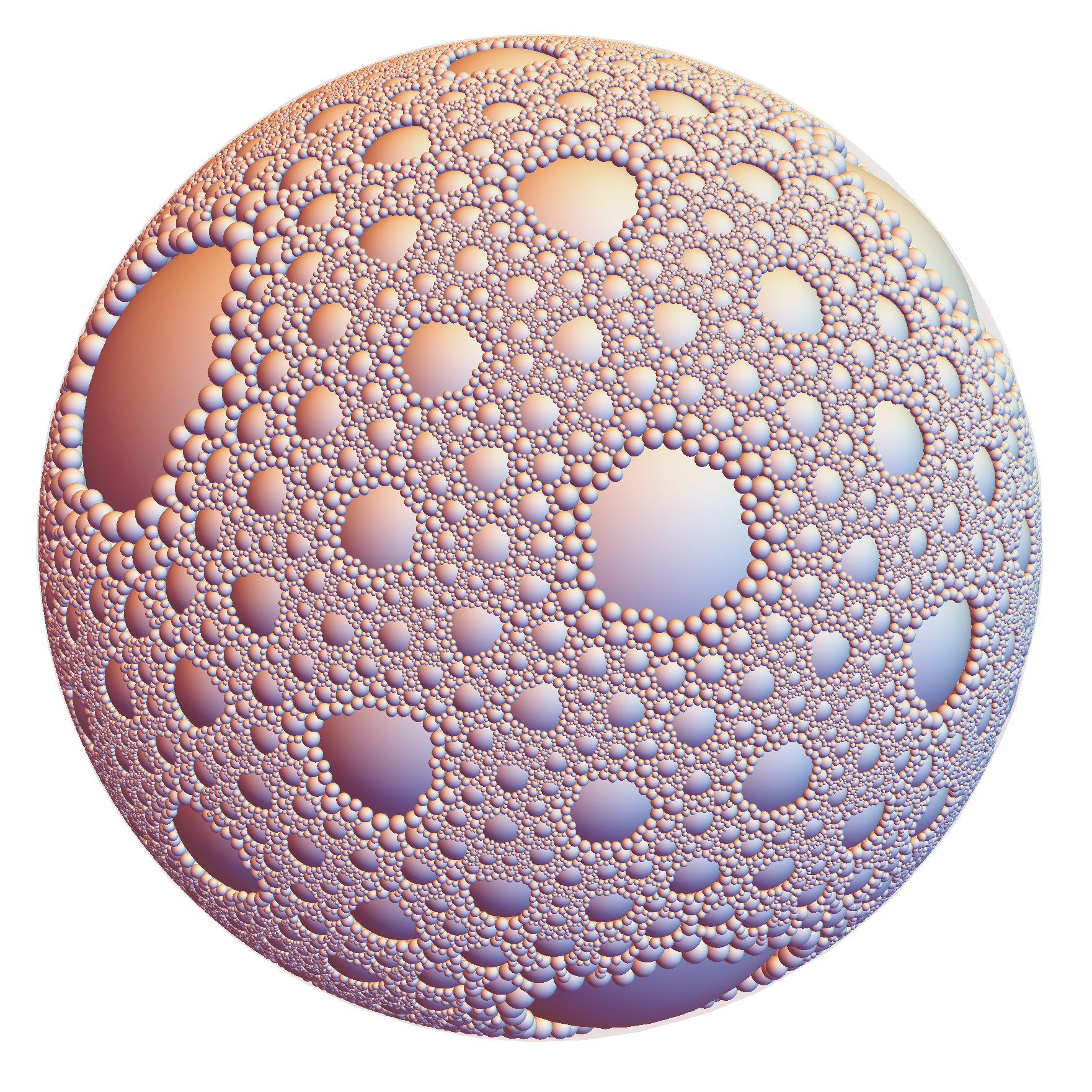}
	\caption{Pentatope-based Apollonian packing with 1\,424\,790 spheres.}
	\label{fig:ApollonianPacking}
\end{figure}

In order to construct an Apollonian packing in two/three 
dimensions, we start with an initial Descartes set of 
four/five pairwise tangent caps on the 2-sphere/3-sphere. 
For this set we calculate the dual caps and determine the 
inversion matrices $ \bi{I} $. We can iteratively generate 
the Apollonian packing if we apply the inversions with 
respect to the initial dual set to all caps generated in 
the previous step, where the zeroth iteration is the 
initial set. This way we fill up the whole 2-sphere. 
However, in three dimensions, we generate several caps 
multiple times due to the overlapping of the dual caps. 
For our purposes and for numerical efficiency, we have 
to remove the duplicates. This we achieve by dividing the 
dual caps into target regions in such a way that each point 
is associated to only one target region. Therefore, we 
construct further caps whose boundaries cross the 
intersection points of the dual caps. New caps are 
accepted only if their centre lies within the target 
region of the inversion. This can easily be tested 
using the scalar product with the dividing caps. 
Remarkably, it is possible to calculate the exact 
positions and sizes of the caps without numerical 
errors since the coordinates take integer values. 
The stereographic projection of the Apollonian packing 
based on the regular pentatope (the four-dimensional 
analogue of the tetrahedron) is shown in 
\fref{fig:ApollonianPacking}.

\begin{figure}
	\centering
	\includegraphics[width=0.6\columnwidth]{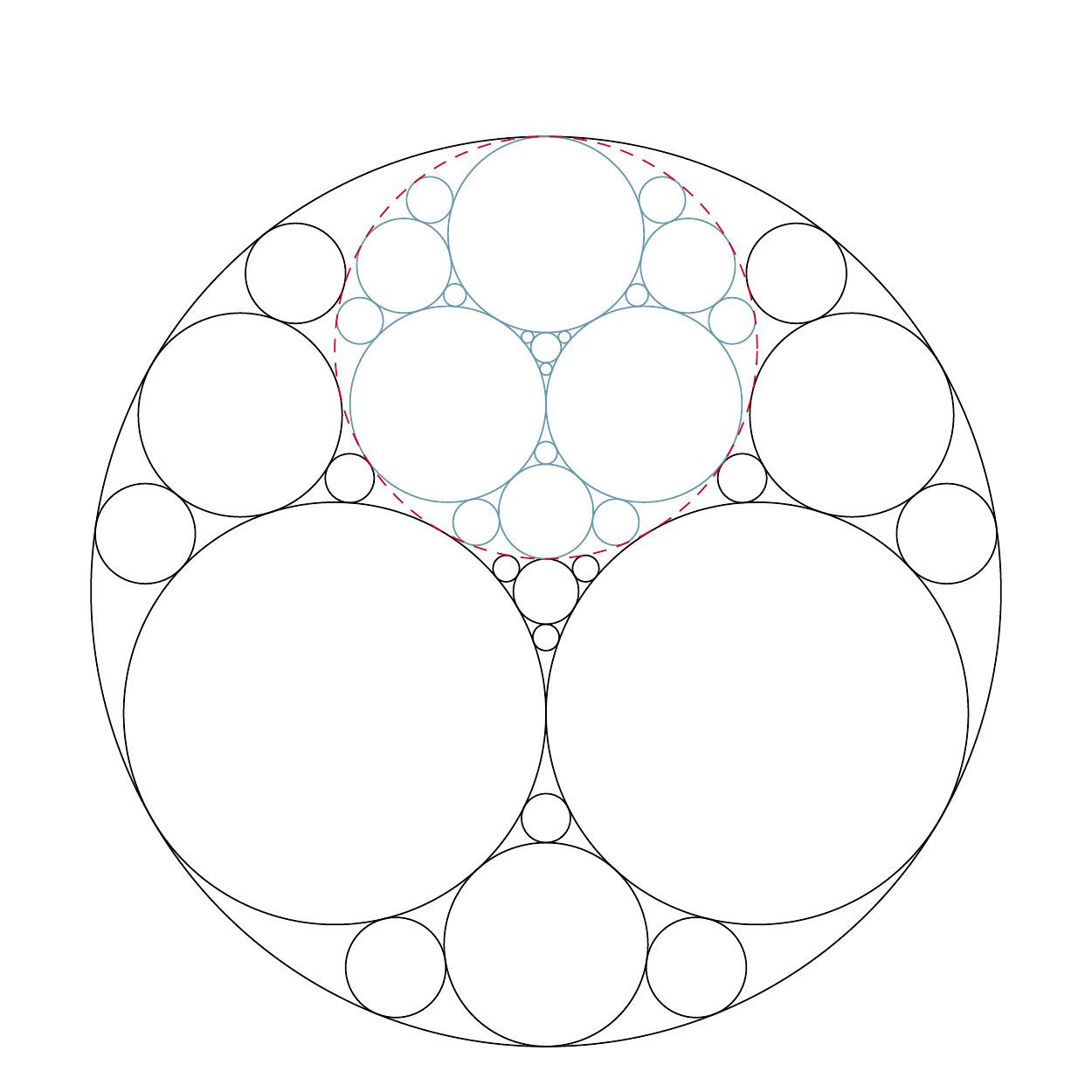}
	\caption{Construction of more uniform packings: A big circle (red) is replaced by a smaller Apollonian packing.}
	\label{fig:UniformPacking}
\end{figure}

In order to obtain more uniform packings without very 
big caps, as we, e.g., want to have for Friedmann-like 
configurations, it is possible to modify this procedure. 
To achieve this, we take the complement of a big cap 
and four new caps inside the former interior of the 
big cap, such that we obtain a new Descartes set. 
Now we repeat the procedure described above and 
generate another Apollonian packing in the former 
interior. In a final step, the complement of the 
original cap is removed. This is shown in 
\fref{fig:UniformPacking}. This procedure can be 
applied to all caps which are too big. Since all 
Apollonian packings are related by a Lorentz 
transformation, it is possible to construct a 
transformation which can be applied to the original 
packing and maps all caps except for one, which 
becomes the exterior, into the interior of a big cap.

\section{Swiss-cheese models}
We already mentioned in the introduction the so called swiss-cheese models for inhomogeneous
cosmologies, the construction of which goes 
back to a seminal paper by Einstein and Straus 
\cite{EinsteinStraus1945}. Their construction 
is based on Friedmann dust universes in which
spherical regions of dust are removed and 
replaced by exterior Schwarzschild geometries. Hence, the global behaviour of such a spacetime 
is still given by the Friedmann equations 
but locally there are regions which are static and 
not influenced by the cosmic expansion. As we 
will use these models for comparison, we want to start by recalling how they are constructed. 
We use units in which $G=c=1$, so that lenghts,
times, and masses share the same unit. We will 
also restrict attention to spherical (positively
curved) dust universes.

A spherical dust universe is described by the 
Friedmann-Lema\^itre-Robertson-Walker metric
\begin{equation}
\label{eqn:FlrwMetric}
	\bi{g} = - \mathbf{d} t^2 + a^2(t) \left( \mathbf{d} \chi^2 + \sin^2 \chi \,\mathbf{d} \Omega^2 \right)\,. 
\end{equation}
The spatial part is a round 3-sphere with a 
time-dependent radius $a(t)$, called scale 
factor. The latter is determined by the first Friedmann equation, here for $\Lambda=0$,
\begin{equation} 
\label{eqn:Friedmann1}
\frac{\dot{a}^2}{a^2} = \frac{8\pi C}{3 a^3}
- \frac{1}{a^2},
\end{equation}
where $C$ is constant. Moreover, space is filled 
with spatially homogeneous dust, that is, an 
ideal fluid with vanishing pressure, $p\equiv 0$, and density given 
by
\begin{equation} 
\label{eqn:Friedmann2}
	\rho(t) = \frac{C}{a^3(t)}.
\end{equation}
Since the volume of the 3-sphere $V(t)= 2 \pi^2 a^3(t) $ 
is finite, it is possible to define a total mass 
via $M=\rho(t) V(t)=2 \pi^2\rho(t)a^3(t)=2\pi^2 C$ 
which is constant due to \eref{eqn:Friedmann2}. 
The first Friedmann equation \eref{eqn:Friedmann1} 
can be solved and the well-known solution in parametric 
form is given by
\begin{eqnarray}
	a (\eta) = \frac{4 \pi C}{3} (1 - \cos \eta), \\
	t (\eta) = \frac{4 \pi C}{3} (\eta - \sin \eta),
\end{eqnarray}
where $\eta\in (0, 2 \pi)$. Hence, the scale factor 
follows a cycloid. The universe starts with a big 
bang and expands to a maximal size 
$a_0=a(\eta =\pi)=\frac{8 \pi C}{3}$. Then it 
recollapses and finally ends in a big crunch. 
It follows that the total mass is given by
\begin{equation}
\label{eqn:TotalMass}
M_{\rm tot} = \frac{3 \pi}{4} \,a_0.
\end{equation}

We cut out the interior of a sphere centred 
at the north pole in the dust universe with 
areal radius $ R = a(t) \,\chi_0 $, where 
$\chi_0 = \mathrm{const}$. Note that the amount 
of dust within that sphere is independent of 
$t$. We now replace the interior geometry, 
which had been of constant positive curvature,  
by that of an exterior Schwarzschild spacetime
describing a black hole with mass $m$. The 
latter is given by
\begin{equation}
\label{eqn:SchwarzschildMetric}
	\bi{g} = - \left( 1 - \frac{2m}{r} \right) \mathbf{d} T^2 + \left( 1 - \frac{2m}{r} \right)^{-1} \mathbf{d}r^2 + r^2 \,\mathbf{d} \Omega^2.
\end{equation}
In these coordinates, the areal radius is just 
$R=r$. In order for this replacement to result 
in a regular solution to Einstein's equations, 
we have to satisfy the Israel junction conditions 
\cite{Israel1966}. For spherically symmetric
 spacetimes, these  conditions have been shown in
\cite{CarreraGiulini} to be equivalent to the 
equality of some physically intuitive quantities on
both sides of the matching spheres along which the 
two spacetimes are glued together. According to 
\cite{CarreraGiulini} it is, in our case, sufficient 
to check the equality of  
together: 
\begin{enumerate}
\item the areal radius $R$, \label{num:ArealRadius}
\item the Misner-Sharp mass $\mathcal{M}$.
\end{enumerate}
Note that the areal radius $R$ is a function defined 
on any spherically symmetric spacetime, the value of 
which at a given point $p$ is defined to be 
$R(p):=\sqrt{A(p)/4\pi}$, where $A(p)$ is the 
2-dimensional volume of the $SO(3)$ orbit containing 
$p$.\footnote{We recall the definition of spherical 
symmetry: A spacetime is called spherically 
symmetric if it allows for an effective $SO(3)$ 
action by isometries whose generic orbits are 
spacelike 2-spheres.} We note the following general 
expression of the Misner-Sharp mass in terms of the 
areal radius, the latter considered as a smooth 
function on spacetime (assigning to each space-time 
point the 2-dimensional area of the $SO(3)$ orbit 
passing through it)
\begin{equation}
\label{eq:MisnerSharp}
\mathcal{M} = \frac{R}{2} \left( 1- \bi{g}^{-1} \left( \mathbf{d}R, \mathbf{d}R \right) \right)	
\end{equation}
Equality of areal radii just means equality of the surface 
areas of the respective $SO(3)$ orbits that are to be 
identified. Equality of the Misner-Sharp masses then means that 
the norms of the differentials $dR$ on these orbits to be 
pairwise identical same. Now, from \eref{eqn:FlrwMetric} and
\eref{eqn:SchwarzschildMetric} one immediately reads off 
that for the FLRW and Schwarzschild geometry the areal radii
are respectively given by 
\begin{eqnarray}
\label{eq:areal-radii-a}
R_{\rm FLRW}&=&\,a(t)\,\sin\chi\,,\\
\label{eq:areal-radii-b}
R_{\rm Schw}&=&r\,.	
\end{eqnarray}
Using this and the expression \eref{eq:MisnerSharp}
for the Misner-Sharp mass, one immediately deduces 
that for FLRW and Schwarzschild the latter is 
respectively given by 
\begin{eqnarray}
\label{eq:ms-masses-a}
\mathcal{M}_{\rm FLRW}
&=&\frac{1}{2}a(t)\bigl(\dot a^2(t)+1\bigr)\sin^3\chi
=\frac{a_0}{2}\sin^3\chi\,,\\
\label{eq:ms-masses-b}
\mathcal{M}_{\rm Schw}&=&m\,,	
\end{eqnarray}
where in the second equality of the first equation
for $\mathcal{M}_{\rm FLRW}$ we have used 
\eref{eqn:Friedmann1} and that the constant $C$ is related
to the maximal scale factor $a_0$ through $a_0=8\pi C/3$, as 
already seen abov

Equality of \eref{eq:ms-masses-a} and \eref{eq:ms-masses-b}
tells us that if a spherical cap of normalised geodesic radius 
$\chi$ (in units of $a(t)$) is removed from the FLRW universe 
and replaced by a Schwarzschild black hole, the mass of the 
latter is given by 
\begin{equation}
\label{eq:equal-ms-masses-consequence}
m=\frac{1}{2}a_0\sin^3\chi\,.	
\end{equation}
Equality of \eref{eq:areal-radii-a} and \eref{eq:areal-radii-b} 
then tell us that the areal radius of the vacuole without dust,
in which the metric is just \eref{eqn:SchwarzschildMetric},
is 
\begin{equation}
\label{eq:equal-areal-radii-consequences}
r=a(t)\sin\chi
\end{equation}
It is time dependent because its boundary is clearly co-moving with the dust. The geometry inside this co-moving vacuole is strictly 
static

This procedure can be repeated for arbitrarily many black holes,
as long as as the Schwarzschild regions do not overlap. If we 
imagine the dust universe as cheese and the Schwarzschild 
regions as holes therein, the intuitive image of a ``swiss-cheese'' 
becomes obvious. We can now construct general swiss-cheese models 
by generating Apollonian packings as described above. 
Every spherical cap of size $\chi$ is then turned into a 
Schwarzschild cell with a black hole at the centre, whose 
(Misner-Sharp) mass equals that of the removed dust and which 
is hence determined by \eref{eq:equal-ms-masses-consequence}. 
Continuing in this fashion by filling in more and mode non
overlapping spherical caps with static vacuum Schwarzschild 
geometries leaves us with as little dust matter as we please, 
and yet the time evolution outside the vacuoles is still 
exactly as in FLRW. We expect that a proper vacuum solution 
to Einstein's equations should be similar to a corresponding 
swiss-cheese model, which will serves us as a reference model.

\section{Exact vacuum initial data}

We wish to compare the swiss-cheese model with an exact 
vacuum solution with black holes of the same masses at 
the same positions. And, as outlined in the introduction, 
the philosophy behind that is to eventually replace 
inhomogeneous matter distributions by inhomogeneous 
distributions of black holes, in which case the time 
evolution is given by Einstein's \emph{vacuum} equations.
The hope connected with that procedure is to eventually 
achieve significant simplifications in the analytical 
and numerical treatments, even though exact analytic 
time evolutions to the initial data representing many
black-holes are not known.  For the moment we are content 
with the fact that it is possible to analytically construct 
exact initial-data on a spacelike hypersurface of constant 
time representing general multi-black-hole configurations. 

In the 3+1-formulation of general relativity, we consider 
time-evolving tensors on a three-dimensional manifold 
instead of tensors on spacetime. This corresponds to a 
foliation of spacetime by spacelike hypersurfaces and 
tensor fields restricted to these. The fundamental 
fields in this theory are the spatial metric $\bi{h}$ 
and the extrinsic curvature $ \bi{K} $, both of which 
are symmetric, purely covariant (all indices down)  
second-rank tensors. For a general review of this 
formalism we refer to \cite{Giulini:SpringerHandbookSpacetime}.

In general relativity, initial data cannot be chosen 
freely but they have to satisfy the Hamiltonian and the 
momentum constraint, which in vacuum ($T_{\mu\nu}=0$) and vanishing cosmological constant read
\begin{eqnarray}
	\mathcal{R}_{\bi{h}}^2 + K^2 - {K^a}_b {K^b}_a = 0, \label{eqn:HamiltonianConstraint} \\
	\nabla_b {K^b}_a - \nabla_a K = 0\,.\label{eqn:MomentumConstraint}
\end{eqnarray}
Here $ \mathcal{R}_{\bi{h}}$ and $\nabla$ are the
Ricci scalar and Levi-Civita covariant derivative 
with respect to the spatial metric $\bi{h}$, respectively, and $K=h^{ab} K_{ab} $ is the trace 
of $\bi{K}$ with respect to $\bi{h}$. As initial
hypersurface, we take a time-symmetric hypersurface
characterised by the vanishing of the extrinsic 
curvature, $ \bi{K} \equiv 0 $. This corresponds 
to a state in which the black holes are momentarily 
at rest. Such a solution should correspond to a 
dust universe at the moment of maximal expansion, 
when the scale factor becomes $a_0$. For 
time-symmetric initial data, the momentum 
constraint \eref{eqn:MomentumConstraint} is 
satisfied identically and the Hamiltonian constraint 
\eref{eqn:HamiltonianConstraint} reduces to the condition of scalar-flatness for the metric 
$\bi{h}$. To satisfy the latter, we make the 
conformal ansatz 
\begin{equation}
\label{eq:ConformalAnsatz}	
\bi{h} = \Psi^4 \,\tilde{\bi{h}} 
\end{equation}
and read the condition for scalar-flatness as 
condition for $\Psi$, whereas the conformal 
metric $\tilde{\bi{h}}$ remains freely specifiable. 
As will be discussed in more detail below (compare
\eref{eq:Conformal-Laplacian2}), this 
leads to an elliptic differential equation for 
$\Psi$, usually referred to as 
\emph{Lichnerowicz equation}, which in our 
case reads: 
\begin{equation}
\label{eqn:LichnerowiczEquation}
\tilde\Delta\Psi - 
\frac{1}{8} \,\tilde\mathcal{R}\,\Psi 
= 0\,.
\end{equation}
Here $\tilde\Delta=\tilde{h}^{ab} \,\tilde\nabla_a \tilde\nabla_b $ is the Laplacian with respect 
to the conformal metric $\tilde{\bi{h}}$.
In view of the cosmological solution \eref{eqn:FlrwMetric}, 
the conformal metric is chosen to be that of a 
round unit 3-sphere\footnote{Here and in the 
sequel $S^3$ always refers to the \emph{unit}
3-sphere.}
\begin{equation}
\label{eq:MetricUnitThreeSphere}
\tilde{\bi{h}} = \bi{h}_{S^3}=\mathbf{d} \chi^2 + \sin^2 \chi \,\mathbf{d} \Omega^2\,,
\end{equation}
where $(\chi,\theta,\varphi)$ are 3-dimensional 
polar angles and $\mathbf{d} \Omega^2:=\mathbf{d}\theta^2+\sin^2(\theta)\,\mathbf{d}\varphi^2$ is the metric of the round unit 2-sphere $S^2_1$. The Ricci scalar of \eref{eq:MetricUnitThreeSphere}
is given by $\tilde\mathcal{R}=6$ 
so that the Lichnerowicz equation
\eref{eqn:LichnerowiczEquation} simply becomes 
\begin{equation}
\label{eq:LichnerowiczEquation_2}
\tilde{\Delta} \Psi - \frac{3}{4} \Psi = 0\,.
\end{equation}
Remarkably, this differential equation is linear 
so that the set of solutions is a linear space 
and the superposition principle applies. Note also 
that solutions cannot 
be globally regular on $S^3$ and must diverge 
somewhere. (Proof: Multiply 
\eref{eq:LichnerowiczEquation_2} with $\Psi$ and 
integrate over $S^3$. Assuming regularity, the 
integral on the left is shown to be strictly 
negative after integration by parts without boundary 
terms, unless $\Psi\equiv 0$; a contradiction!) 
The non-regular points will be removed 
without introducing any (geodesic- and Cauchy-) 
incompleteness in the manifold 
$S^3-\{\mathrm{non\ regular\ points}\}$ with 
Riemannian metric $\bi{h}$. This is because 
the diverging $\Psi$ will send the non-regular
points to an infinite distance with respect to 
the metric $\bi{h}=\Psi^4\tilde\bi{h}$. After point 
excision, the remaining neighbourhood of each 
point is an asymptotically flat end of the 
initial-data 3-maifold and represents a black 
hole. 

\subsection{Time symmetric multi black-hole
solutions to Lichnerowicz equation}
Linearity allows to give solutions to \eref{eq:LichnerowiczEquation_2} for an 
arbitrary number of black holes. They are 
easily written down if we think of the unit 
$S^3$ embedded in euclidean $\mathbb{R}^4$. 
If we write $\bi{X}$ for the point of the 
3-sphere (which one may think of as being 
parametrised by, say, the polar angles 
$(\chi,\vartheta,\varphi)$ or, alternatively, 
Euler angles $(\psi,\vartheta,\varphi)$, if 
one prefers to think in terms of coordinates, though we will not make 
use of such coordinatisations) 
and $\Vert \,\cdot\, \Vert $ for the standard 
(euclidean) norm of $ \mathbb{R}^4 $, the
solution for a number $N$ of black holes 
is then given by
\begin{equation}
\label{eq:Solution-LichnerowiczEquation_2}
\Psi (\bi{X}) = \sum_{i=1}^{N} \frac{\mu_i}{\| \bi{X} - \bi{P}_i \|}.
\end{equation}
The solution property for each of the $N$ terms 
is proven in detail in \ref{eq:Appendix-1}, as a 
special case of a more general theorem that works 
in all dimensions. 

The point $\bi{P}_i\in S^3$ corresponds to the 
``position'' of the $i$-th black hole and the 
parameters $\mu_i$ are related to the masses by
the expressions 
\begin{equation}
\label{eq:ADM-Mass-general}
m_i
=2\sum_{j=1\atop j\ne i}^{N}
\frac{\mu_j\mu_i}{\Vert\bi{P}_j-\bi{P}_i\Vert}
\qquad(1\leq i\leq N)
\,,
\end{equation}
which we will derive below. The $N$ points 
$\bi{P}_i$ where the solution diverges are 
removed from the manifold without introducing 
any incompletenesses.
In fact, for $\bi{X}\rightarrow\bi{P}_i$ the 
metric is asymptotically flat and we will refer 
to this region as an ``end''.%
\footnote{The notion of ``end'' for a topological 
space was introduced by Freudenthal
\cite{Freudenthal:1931}. Roughly speaking, an end 
is a connected component in the complement of arbitrarily large compact sets.}
Topologically the manifold is the $N$-fold 
punctured $S^3$. This solution is also 
discussed in \cite{Clifton.EtAl:2012} and
\cite{Bentivegna.EtAl:2018} in slightly different
but equivalent presentations. Our presentation 
\eref{eq:Solution-LichnerowiczEquation_2} makes 
use of the simple embedding geometry of 
$\mathbb{R}^4$, which leads to simpler 
expressions and is much better adapted to
later applications of Lie sphere-geometry. 
But for completeness and comparison we note that 
the $\mathbb{R}^4$-distance $\Vert\bi{X}-\bi{Y}\Vert$
and the intrinsic geodesic distance (compare \eref{eq:DefGeodDistance}) 
$\Lambda=\Lambda(\bi{X},\bi{Y}):=\arccos(\bi{X}\cdot\bi{Y})$
between two points $\bi{X}$ and $\bi{Y}$ on $S^3$
are simply related by $\Vert\bi{X}-\bi{Y}\Vert=\sqrt{2(1-\cos(\Lambda))}=2\,\sin(\Lambda/2)$.
This is the way the solution was recently 
presented and discussed in 
\cite{Bentivegna.EtAl:2018,Durk.Clifton:2017},
with generalisation to non-vanishing cosmological 
constant in \cite{Durk.Clifton:2017b}.
\begin{figure}[h!]
	\centering
	\includegraphics[width=0.5\columnwidth]{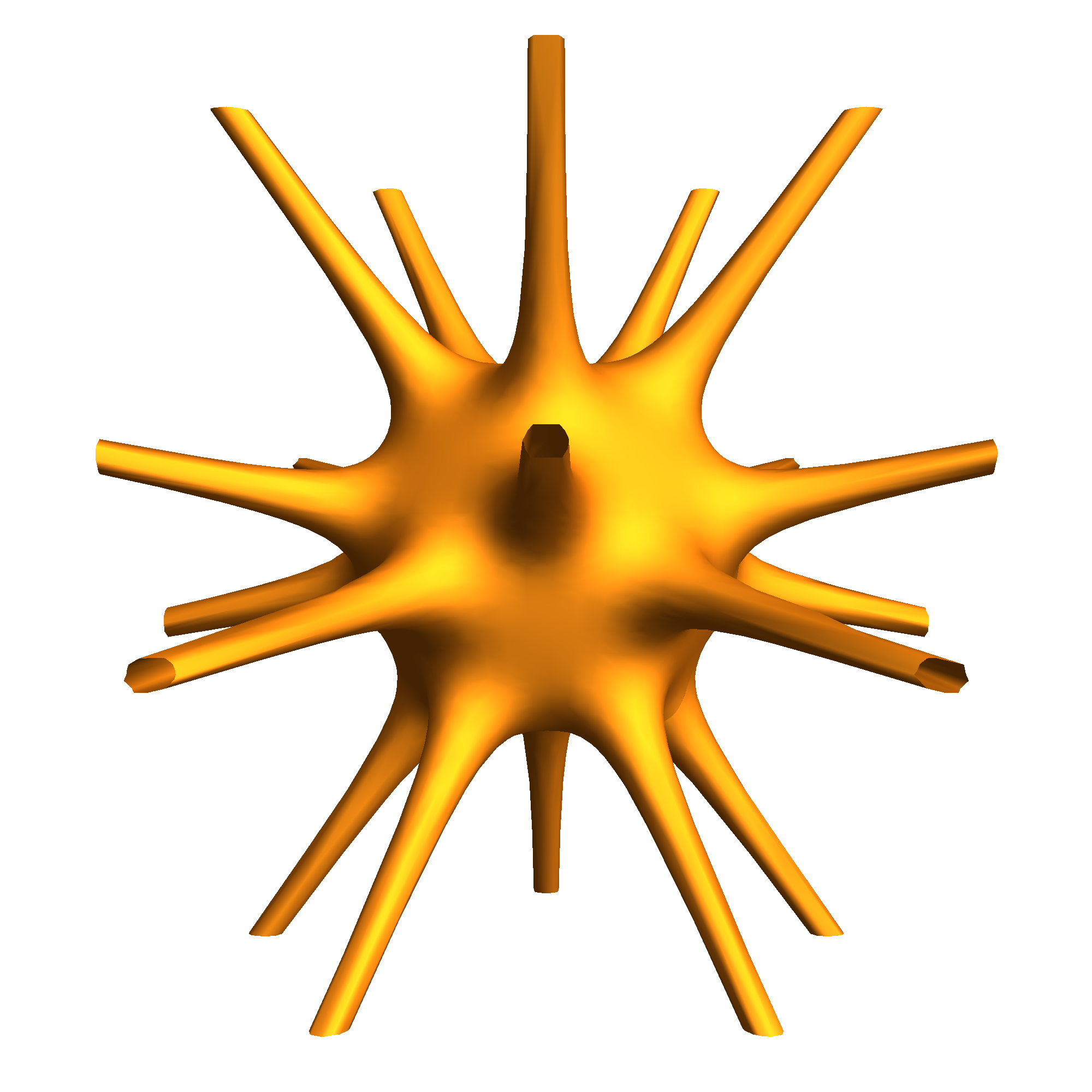}
	\caption{Plot of the function $\Psi$ given
in \eref{eq:Solution-LichnerowiczEquation_2} 
over the 3-sphere, here represented as 2-sphere.
It diverges at the $N$ poles $\bi{P}_i$ which 
are deleted from the manifold. Neighbourhoods
of the deleted points where $\Psi$ is large 
then correspond to asymptotically flat ends, 
of which there are $N$, and which are 
geodesically complete. 
\label{fig:ConformalFactor}}
\end{figure}

\subsection{Isometry to Brill-Lindquist data}
It is instructive to note that the solution just 
found is just the same (i.e. isometric to) as 
the good old Brill-Lindquist initial data sets \cite{Brill.Lindquist:1963} for $(N-1)$ 
black holes in an asymptotically flat 3-manifold the 
topology of which is that of a $(N-1)$-fold punctured 
$\mathbb{R}^3$. In fact, there are $N$ isometries of 
our solution to such Brill-Lindquist sets, given 
by the stereographic projections 
$\pi:S^3-\{\bi{P}\}\rightarrow\mathbb{R}^3$, where the 
pole $\bi{P}$ of the projection is chosen to be any 
of our black-hole positions $\bi{P}_i$, followed by a constant rescaling $\bi{x}\mapsto\bi{x}':=(\mu_i^2/2)\,\bi{x}$. 

Before writing out the details of this isometry, 
let us point out that its existence is obvious 
from the conformal properties of the Laplacian
and the conformal flatness of the metric $\bi{h}
_{S^3}$ of the unit 3-sphere, expressed in 
formula \eref{eq:Appendix2-9} of \ref{sec:Appendix2}.
Quite generally, the following is true (see, e.g.,
\cite{Giulini:SpringerHandbookSpacetime} for proofs 
and further details): Let $(M,\bi{g})$ be a 
(Semi-) Riemannian manifold of dimension $n>2$ 
and consider on $C^\infty(M,\mathbb{R})$ the 
$\bi{g}$-dependent linear differential operator 
(sometimes called the ``conformal Laplacian'')
\begin{equation}
\label{eq:Conformal-Laplacian1}
\mathcal{D}_{\bi{g}}:=
\Delta_{\bi{g}}-
\frac{n-2}{4(n-a)}\mathcal{R}_{\bi{g}}\,,
\end{equation}
where $\Delta_{\bi{g}}$ and $\mathcal{R}_{\bi{g}}$ denote the Laplacian and Ricci scalar with 
respect to $\bi{g}$, respectively. Let $M_{\Omega}$ denote the linear operator in 
$C^\infty(M,\mathbb{R})$ that multiplies each 
element with $\Omega\in C^\infty(M,\mathbb{R}_+)$.
Then the following relation holds: 
\begin{equation}
\label{eq:Conformal-Laplacian2}
\mathcal{D}_{\Omega^{\frac{4}{n-2}}\bi{g}}
=M_{\Omega^{-\frac{n+2}{n-2}}}
\circ\mathcal{D}_{\bi{g}}
\circ M_{\Omega}\,.
\end{equation}
In $n=3$ dimensions we have 
$\mathcal{D}_{\bi{g}}=\Delta_g-(1/8)\mathcal{R}_g$. 
Equation \eref{eq:Conformal-Laplacian2} and 
conformal flatness\footnote{Here and in the sequel 
$\mathbb{R}^3$ denotes flat euclidean 3-space 
endowed with with its natural coordinates $x^a$ 
in which the flat metric is 
$\bi{h}_{\mathbb{R}^3}=
\sum_{a=1}^3\mathbf{d}x^a\otimes\mathbf{d}x^a=\mathbf{d}\bi{x}\dot\otimes\mathbf{d}\bi{x}$.}
 of the unit-sphere metric, 
i.e., $\bi{h}_{S^3}=\Omega^4\bi{h}_{\mathbb{R}^3}$,
immediately imply that if $\Psi$ is in the kernel 
of $\mathcal{D}_{\bi{h}_{S^3}}$, i.e. solves  \eref{eq:LichnerowiczEquation_2},then 
$\Omega\cdot\Psi$ is in the kernel of 
$\mathcal{D}_{\bi{h}_{\mathbb{R}^n}}=
\Delta_{\bi{h}_{\mathbb{R}^n}}$ and hence harmonic. The latter are the 
solutions to the Lichnerowicz equation in the conformally flat Brill-Lindquist case, the former 
are our solutions in the conformally spherical cosmological case. hence we see that they are just related by multiplication with (a constant multiple
of) $\Omega$. This we will now show more explicitly.   

We are interested in the explicit form of this isomorphism, for that will provide analytic 
expressions relating the parameters $\mu_i$ with the familiar expressions for the ADM-masses of the 
black holes. For the reader's convenience we have
collected the relevant facts and formulae 
concerning stereographic projections and its 
metric properties in \ref{sec:Appendix2} in an 
essentially  coordinate independent form. Given
these formulae, the explicit proof of isometric equivalence is easy. We write \eref{eq:ConformalAnsatz} with $\tilde\bi{h}=\bi{h}_{S^3}$,
replace $\bi{h}_{S^3}$ according to \eref{eq:Appendix2-9}
with the flat metric $\bi{h}_{\mathbb{R}^3}$ and replace 
$\Psi$ with the right-hand side of \eref{eq:Solution-LichnerowiczEquation_2}; this gives: 
\begin{equation}
\label{eq:BrillLindquist-Isom1}	
\bi{h}=\left(\sum_{i=1}^N\frac{\mu_i}{\Vert\bi{X}-\bi{P}_i\Vert}\right)^4
\frac{\Vert\bi{X}-\bi{P}\Vert^4}{4}\,\bi{h}_{\mathbb{R}^3}\,.
\end{equation}
Now we choose any of the black-hole ``positions'' $\bi{P}_i$
as center $\bi{P}$ for the stereographic projection, say 
$\bi{P}=\bi{P}_N$. Then 
\begin{equation}
\label{eq:BrillLindquist-Isom2}	
\bi{h}=\left(1+\sum_{i=1}^{N-1}\frac{\mu_i}{\mu_N}\frac{\Vert\bi{X}-\bi{P}_N\Vert}{\Vert\bi{X}-\bi{P}_i\Vert}\right)^4
\frac{\mu^4_N}{4}\,\mathbf{d}\bi{x}\dot\otimes \mathbf{d}\bi{x}\,.
\end{equation}
Setting $\bi{P}=\bi{P}_N$ and $\bi{Y}=\bi{P}_i$ in 
equation \eref{eq:Appendix2-6} of \ref{sec:Appendix2} 
shows that   
\begin{equation}
\label{eq:BrillLindquist-Isom3}	
\frac{\Vert\bi{X}-\bi{P}_N\Vert}%
{\Vert\bi{X}-\bi{P}_i\Vert}
=\frac{2}{\Vert\bi{P}_i-\bi{P}_N\Vert}
\cdot
\frac{1}{\Vert\bi{x}-\bi{p}_i\Vert}\,,
\end{equation}
where $\bi{x}$ and $\bi{p}_i$ are the images of 
$\bi{X}$ and $\bi{P}_i$ under the stereographic 
projection. Hence \eref{eq:BrillLindquist-Isom2} 
can be rewritten into  	
\begin{equation}
\label{eq:BrillLindquist-Isom4}	
\bi{h}=\left(1+\sum_{i=1}^{N-1}\frac{\lambda_i}%
{\Vert\bi{x'}-\bi{p}'_i\Vert}\right)^4
\,\mathbf{d}\bi{x}'\dot\otimes \mathbf{d}\bi{x}'\,,
\end{equation}
where $\bi{x}':=(\mu^2_N/2)\,\bi{x}$,
$\bi{p}_i':=(\mu^2_N/2)\,\bi{p}_i$, and
\begin{equation}
\label{eq:BrillLindquist-Isom5}
\lambda_i:=\frac{\mu_i\mu_N}{\Vert\bi{P}_i-\bi{P}_N\Vert}\,.
\end{equation}
Equation \eref{eq:BrillLindquist-Isom5} are 
precisely the Brill-Lindquist data for $(N-1)$ 
black holes at positions 
$\bi{p}'_i=(\mu_N/2)\,\pi(\bi{P}_i)$. The 
manifold is $\Sigma:=\mathbb{R}^3-\{\bi{p}'_1,\cdots,\bi{p}'_{N-1}\}$ with coordinates $\bi{x}'$ with 
respect to which the initial metric is the 
canonical flat metric $\,\mathbf{d}\bi{x}'\cdot \mathbf{d}\bi{x}'$. 
The Riemannian manifold $(\Sigma,\bi{h})$ is 
complete with $N$ asymptotically flat ends, one 
for $\Vert\bi{x'}\Vert\rightarrow\infty$ (spacelike
infinity) and $(N-1)$ ``internal'' ones, one for 
each $\bi{x'}\rightarrow\bi{p}'_i$, where $i=1,\cdots, (N-1)$. 

\subsection{ADM masses}
Quite generally, an ADM mass can be associated to 
any asymptotically flat end of a 3-manifold 
in a purely geometric fashion \cite{Bartnik1986};
for applications compare also \cite{Giulini:SpringerHandbookSpacetime}). 
The invariant geometric character of this 
association allows to compute the ADM mass in 
suitable coordinates. A convenient way to do this 
is to asymptotically put the metric towards the 
flat end into the form of the spatial part of the 
exterior Schwarzschild metric in so-called 
isotropic coordinated (which also manifestly
display conformal flatness). Then the metric 
takes the form
\begin{equation}
	\label{eq:SpatialSchwarzschild}
\bi{h}_{\rm Schw}=\left(1+\frac{m}{2r}\right)^4\,
(\mathbf{d}r\otimes\mathbf{d}r+r^2\,\bi{h}_{S^2})\,,
\end{equation}
where $m$ is the ADM-mass in geometric units 
(i.e. $m=GM/c^2$, where $M$ is the mass in SI-units)
and $\bi{h}_{S^2}$ denotes the standard round metric
on the unit 2-sphere. 

In our case, there is one such ADM mass for each 
of the $N$ ends of $(\Sigma,\bi{h})$. That at spatial 
infinity we call $m_N$, for on $S^3$ it corresponds to
the black hole at $\bi{P}_N$. Here, in the 
Brill-Lindquist picture, it corresponds to the total 
mass/energy of spacetime, that is composed of all 
the contributions of all $(N-1)$ black holes, 
diminished by the (negative) binding energy (compare 
the discussions in \cite{Brill.Lindquist:1963} and \cite{Giulini:SpringerHandbookSpacetime}). 
Direct comparison of \eref{eq:BrillLindquist-Isom4} 
for $\Vert\bi{x'}\Vert\rightarrow\infty$ with \eref{eq:SpatialSchwarzschild} immediately gives
\begin{equation}
	\label{eq:ADM-Mass-N}
m_N
=2\sum_{i=1}^{N-1}\lambda_i
=2\sum_{i=1}^{N-1}
\frac{\mu_i\mu_N}{\Vert\bi{P}_i-\bi{P}_N\Vert}
\end{equation}
The other masses can also be directly computed 
within the same stereographic projection, as we 
will show next. However, we can, in fact, 
immediately tell the result without any further calculation. This is true because we could have 
chosen any of the points $\bi{P}_j$ as centre 
for the stereographic projection, which would 
have resulted in the corresponding formula to \eref{eq:ADM-Mass-N}, with $j$, rather than $N$, 
being the distinguished index. This indeed just 
leads to \eref{eq:ADM-Mass-general}.

Despite this latter argument is elegant and 
certainly correct, we still wish to show how 
one arrives at the same result within the same 
stereographic projection centred at $\bi{P}_N$.
The reason is that this calculation is 
instructive insofar as it shows how a well 
known expression for black-hole masses in 
the conformally flat Brill-Lindquist approach 
are rendered much more symmetric in the 
conformally spherical cosmological approach 
discussed here. The direct calculation proceeds 
as follows: For any 
$1\leq i\leq (N-1)$ choose ``inverted'' spherical 
polar coordinates $(\rho_i,\theta,\varphi)$ based 
at $\bi{p}'_i$, where 
$\rho_i:=\lambda_i^2/\Vert\bi{x}'-\bi{p'_i}\Vert$. 
The limit $\bi{x}'\rightarrow\bi{p}_i$ then 
corresponds to $\rho_i\rightarrow\infty$. 
In these coordinates the metric then assumes the 
form  \eref{eq:SpatialSchwarzschild} with 
$r=\rho_i$ and 
\begin{equation}
\label{eq:Brill-Lindquist-Masses}
m=m_i:=2\lambda_i\,\biggl(1+\sum_{j\ne i}\lambda_j/\Vert\bi{p}'_j-\bi{p}'_i\Vert\biggr)
\qquad (1\leq i\leq (N-1))\,.
\end{equation}
This formula for the mass of a single hole in the 
metric \eref{eq:BrillLindquist-Isom4} is well known
from \cite{Brill.Lindquist:1963}. Now, replacing 
all $\lambda_i$ according to 
\eref{eq:BrillLindquist-Isom5}, setting 
 $\Vert\bi{p}'_j-\bi{p}'_i\Vert=\mu^2_N/2\Vert\bi{p}_j-\bi{p}_i\Vert$ and replacing $\Vert\bi{p}_j-\bi{p}_i\Vert$ by means of \eref{eq:Appendix2-6} with 
$\bi{x}=\bi{p_i}$, $\bi{y}=\bi{p_j}$, and 
$\bi{P}=\bi{P}_N$ then gives indeed
\eref{eq:ADM-Mass-general}. Note that the $(N-1)$
expressions \eref{eq:Brill-Lindquist-Masses} for 
the individual holes all look the same, but clearly
different from the expression given by the first
equality in \eref{eq:ADM-Mass-N} for the overall 
energy of all $(N-1)$ holes taken together, 
whereas in the conformally spherical cosmological picture the $(N-1)+1=N$ expressions 
\eref{eq:ADM-Mass-general} are again symmetric.  

\subsection{Geometry and topology}
Finally we wish to mention a few more aspects in connection with the geometry and topology of the 
initial-data surface $\Sigma:=\mathbb{R}^3-\{\bi{p}'_1,\cdots,\bi{p}'_{N-1}\}$ in the Brill-Lindquist picture. Its geometry is conformally 
flat, $\bi{h}=\Psi^4\bi{h}_{\mathbb{R}^3}$, where 
$\Psi$ satisfies Laplace's equation 
$\Delta_{\mathbb{R}^3}\Psi=0$, which is what 
Lichnerowicz's equation reduces to in this 
case. The solution given in 
\eref{eq:BrillLindquist-Isom4}, i.e.
\begin{equation}
\label{eq:BrillLindquistPsi}
\Psi(\bi{x}')=1+\sum_{i=1}^{N-1}\frac{\lambda_i}{\Vert\bi{x}'-\bi{p}'_i\Vert}\,,	
\end{equation}
is essentially a sum of $(N-1)$ monopoles without 
contributions from higher multipoles. One might 
wonder why higher multipoles were excluded. 
The answer is 
that any such higher multipole would render the 
metric $\bi{h}$ incomplete ($\Psi$ acquires zeros). Without higher multipoles, each monopole renders 
the manifold asymptotically flat in a neighbourhood of its location $\bi{p}'_i$ and introduces one end 
to which an ADM mass can be associated. Also 
associated to each end is an outermost (as seen 
from the end) minimal surface which, since we 
consider time-symmetric initial data, is an 
apparent horizon. In that sense the initial 
data set contains $(N-1)$ black holes.
Note also that $\Sigma$ is connected and simply 
connected, but with non-trivial second homology
group given by 
\begin{equation}
\label{eq:SecondHomology}
H_2(\Sigma,\mathbb{Z})=\mathbb{Z}^{N-1}	
\end{equation}
which in this case (i.e. due to simple
connectedness) is also isomorphic to the second
homotopy group $\pi_2(\Sigma)$. 
Each of the $(N-1)$ factors 
$\mathbb{Z}$ in  \eref{eq:SecondHomology} is 
generated by one of the apparent horizons. 
There may be additional minimal surfaces 
corresponding to other elements of  \eref{eq:SecondHomology}, like the sums 
of generators, which enclose the corresponding 
set of black holes if their positions are 
chosen sufficiently close together (the individual 
holes may then be said to have merges into a 
composite black hole). In the extreme case, where 
\emph{all} the $(N-1)$ holes are sufficiently close,
there will be an $N$th minimal surfaces enclosing
all of them and corresponding to the sum all all 
generators in \eref{eq:SecondHomology}. This is the 
situation we have in mind if we speak of $N$ 
black holes on the 3-sphere. But note that in our 
original conformally spherical picture, adding just 
a single pole results in flat space without 
any black hole and adding two poles merely results 
in the outer Schwarzschild geometry representing 
a single hole. For $N>2$ poles the data 
result in at least $N-1$ black holes, and 
possibly $N$ if the data are suitably chosen.

Finally we remark that the solution corresponding 
to the swiss-cheese model is obtained if we take 
the centres of the spherical caps for $\bi{P}_i$ 
and the mass parameters are obtained by solving the coupled system \eref{eq:ADM-Mass-general}
of quadratically equations for $\mu_i$ which can 
be done only numerically. 

\section{Unifoamy configurations}
\label{sec:UnifoamyConfigurations}
\begin{figure}[h!]
\centering
\includegraphics[width=0.5\columnwidth]{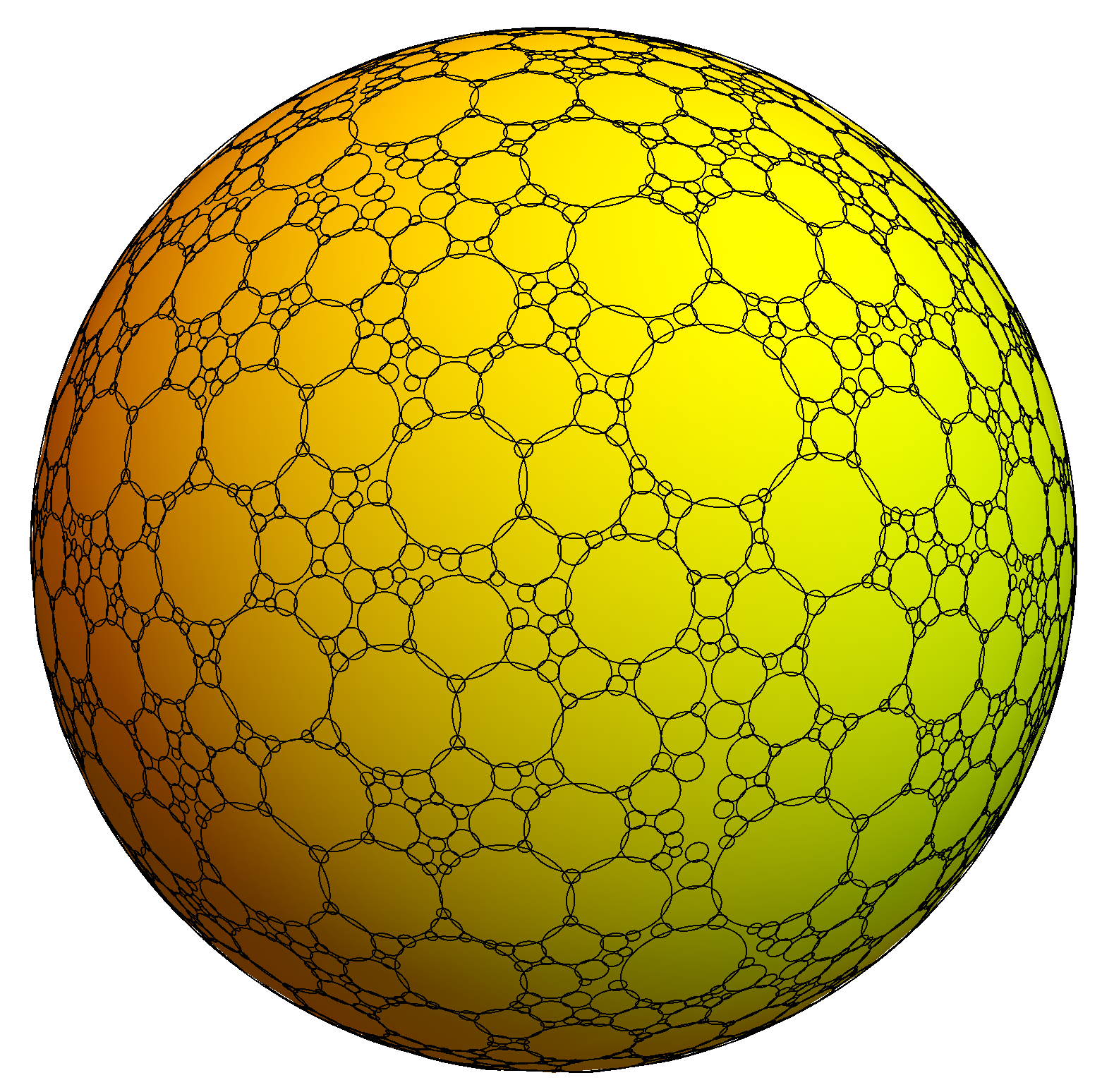}
\caption{%
Two-dimensional illustration of the Lindquist-Wheeler
model of a unifoamy configuration (the central 
black holes are not plotted). The cells are distributed 
quite evenly on the sphere, they are not too big 
and do not overlap too much. Since this pictures 
gives the impression of a uniform foam on a sphere, 
we called such configurations ``unifoamy''.
\label{fig:Unifoamy}}
\end{figure}

We have two solutions with Schwarzschild(-like) black holes of the same masses at the same positions: the swiss-cheese model at the moment of maximal expansion and the initial data. Which Friedmann dust universe approximates such a solution best? In the former case, we simply take the dust universe of the model. In the latter case, we expect a similar value if most of the dust in the corresponding swiss-cheese model is removed. Clearly, not every configuration of black holes resembles a Friedmann dust universe. Therefore, the black holes should be distributed somehow evenly on the 3-sphere. However, there is no general notion on a uniform distribution of points on the 3-sphere and the definition of uniformity depends on the problem. Our approach is as follows. The mean inverse distance between two points in a uniform density distribution, $ \rho = \mathrm{const} $, is given by
\begin{equation}\fl
	\left\langle \frac{1}{\| \bi{P}_i - \bi{P}_j \|} \right\rangle = \frac{1}{2 \pi^2} \int_{0}^{\pi} \rmd \chi \int_{0}^{\pi} \rmd \vartheta \int_{0}^{2 \pi} \rmd \varphi \sin^2 \chi \sin \vartheta \,\frac{1}{\sqrt{2 (1 - \cos \chi)}} = \frac{8}{3 \pi},
\end{equation}
using a coordinate system such that one point is located at the north pole. For a discrete configuration of equal black holes, we simply demand the discrete analogue, namely
\begin{equation}
	\left\langle \frac{1}{\| \bi{P}_i - \bi{P}_j \|} \right\rangle = \frac{1}{N} \sum_{j \neq i} \frac{1}{\| \bi{P}_i - \bi{P}_j \|} = \frac{8}{3 \pi}
\end{equation}
for all points $ \bi{P}_i $. In the general case, we weight the inverse distances with the mass parameters, yielding
\begin{equation}
	\left\langle \frac{1}{\| \bi{P}_i - \bi{P}_j \|} \right\rangle = \frac{1}{\sum_{k \neq i} \mu_k} \sum_{j \neq i} \frac{\mu_j}{\| \bi{P}_i - \bi{P}_j \|} = \frac{8}{3 \pi}.
\end{equation}
If we multiply this equation with $ 2 \mu_i $, we obtain after a rearrangement
\begin{equation} 
\label{eqn:UnifoamyMass}
	m_i = \sum_{j \neq i} \frac{2 \mu_i \mu_j}{\| \bi{P}_i - \bi{P}_j \|} = \frac{16}{3 \pi} \sum_{j \neq i} \mu_i \mu_j.
\end{equation}
Hence, our condition for Friedmann-like configurations constrains the mass of each black hole which is now essentially determined by its mass parameter irrespectively of the positions of all other black holes on the 3-sphere in this case. This condition also guarantees that the black holes are not too close to each other. We call configurations satisfying \eref{eqn:UnifoamyMass} \emph{unifoamy} since it seems that the corresponding swiss-cheese model consists of evenly distributed Schwarzschild cells or, illustratively, a \emph{uniform foam} of 
Schwarzschild bubbles. This is illustrated in \fref{fig:Unifoamy}.
In passing we note that unifoamy configurations 
can be related to \emph{central configurations};
compare \cite{BattyeEtAl2003} for the general 
notion and \cite{EllisGibbons2013} for 
applications to Newtonian cosmology.  
Central configurations come into play if, for a 
fixed set of parameters $\mu_i$, we ask for the 
set of positions $\bi{P}_i$ on $S^3$ for which 
the sum of masses $m_i$ according to 
\eref{eqn:UnifoamyMass}, i.e. the function 
$\sum_i\sum_{j\ne i}2\mu_i\mu_j/
\Vert\bi{P}_i-\bi{P}_j\Vert$, takes its 
minimal value. Adding the $N$ constraints 
$\bi{P}^2_i-1=0$ with $N$ Lagrange multipliers 
$\lambda_i$ and carrying out the variation with 
respect to each position  $\bi{P}_i$ and each 
multiplier $\lambda_i$ results in equations 
which for $\lambda_i=C\mu_i$ turn into the 
equations for central configurations~\cite{Fennen2017}.

In order to be similar to a spherical Friedmann dust universe, we have to fit two parameters: the size $ a_0 $ and the total mass $ M $. We set the total mass of the black holes to $ M = \sum_i m_i $. Since the size and the total mass of a spherical dust universe are related by \eref{eqn:TotalMass}, it appears to be natural to take this as the definition of the fitted size. Hence, we obtain for the size
\begin{equation}\label{eqn:UnifoamySize}
	A_0 = \frac{4}{3 \pi} M = \frac{4}{3 \pi} \sum_i m_i = \frac{64}{9 \pi^2} \sum_i \sum_{j \neq i} \mu_i \mu_j.
\end{equation}
For this reason, the total mass automatically fits to the dust universe and we only have to argue that our choice $ A_0 $ for the size also fits. This means that, comparing the spatial metric of a dust universe with the one of the black hole initial data, the deviation of $ \Psi^2 $ from the fitted size $ A_0 $ should be small in the far-field region of the black holes. Clearly, the deviation is large in the vicinity of the black holes. We are not expecting that the space resembles a dust universe close to a black hole in correspondence with our Universe in which local dynamics in the regime of galaxies strongly differ from the behaviour of the Universe on cosmic scales. 

For a large number of black holes, our result is approximately 
the same which was obtained by Korzy\'nski by an ad-hoc averaging 
procedure \cite{Korzynski2014}. He averaged the conformal factor 
$\Psi$ over the 3-sphere with respect to the round metric, yielding
\begin{equation}
	\langle \Psi \rangle = \frac{1}{2 \pi^2} \int_{S^3} \Psi \,\rmd V = \frac{64}{9 \pi^2} \sum_{i,j} \mu_i \mu_j \approx A_0.
\end{equation}
Korzy\'nski could give upper bounds for the deviation of the conformal factor from its average. The main parameters are the distance to the closest black hole with respect to the round metric and the so-called modified spherical cap discrepancy $ \mathcal{E} $ which is a quite abstract object and difficult to compute for a particular configuration. However, for particular configurations it is possible to estimate the cap discrepancy as follows. If we divide the 3-sphere in non-overlapping regions $ \mathcal{V}_i $ such that the whole 3-sphere is covered and each region contains a black hole whose mass parameter is proportional to the volume of the region, $ \mu_i = \kappa \, \mathrm{vol} \,\mathcal{V}_i $, the spherical cap discrepancy is bounded from above by the largest diameter of all regions, that is,
\begin{equation}
	\mathcal{E} \leq \max_{i=1,\dots,N} \mathrm{diam} \,\mathcal{V}_i,
\end{equation}
where $ \mathrm{diam} \,\mathcal{V}_i = \sup_{\bi{X}, \bi{Y} \in \mathcal{V}_i} \Lambda (\bi{X}, \bi{Y}) $. If we consider a configuration which is generated by an Apollonian packing, it should be possible to slightly deform the spherical caps such that the estimate is still approximately valid and given my the largest size $ \chi_0 $ of all spherical caps, 
\begin{equation}
	\mathcal{E} \lesssim 2 \chi_0.
\end{equation}
The mass of the black holes is related to the size of the spherical cap by 
\eref{eq:equal-ms-masses-consequence}.
 If we substitute the size $a_0$ by the total mass 
according to \eref{eqn:TotalMass}, now simply 
writing $M$ instead of $M_{\rm tot}$,
and solve for the size $\chi$, we obtain
\begin{equation}
	\chi = \arcsin \left[ \left( \frac{3 \pi}{2 M} \,m_i \right)^{1/3} \right].
\end{equation}
Hence, a good estimate for the spherical cap discrepancy should be given by
\begin{equation}
	\mathcal{E} \lesssim 2 \kappa \arcsin \left[ \left( \frac{3 \pi}{2 M} \max m_i \right)^{1/3} \right].
\end{equation}
Therefore, we expect for configurations of black holes with similar masses, that the deviation of the conformal factor from its average decreases in most regions because the cap discrepancy decreases with an increasing number of black holes in this case. Since $ \langle \Psi \rangle^2 \approx A_0 $, the same should hold for our fit $ A_0 $. Hence, the space is almost round as it should be for a spherical dust universe. In particular, the minimum of the conformal factor $ \Psi_{\mathrm{min}} $, which is taken in the far field of the black holes, should be close to averaged value and therefore $ \Psi_{\mathrm{min}}^2 \approx A_0 $.

\section{Comparison and Discussion}

Finally, we want to compare the different fits 
for the size to initial data configurations 
with the corresponding reference model. 
By this we mean the swiss-cheese model with 
black holes located at the same positions and
endowed with the same masses. For a good 
approximation, we expect that the fitted size 
is close to the size of the reference model, 
that is, the radius $ a_0 $ of the dust universe 
in the swiss-cheese model.

We consider the configurations with black holes on 
the centres $ \bi{P}_i $ of the spheres in the 
Apollonian packings as presented above. The masses 
$m_i$ of the black holes are given by the opening 
angles $ \alpha_i $ of the spherical caps via 
\eref{eq:equal-ms-masses-consequence}. 
The mass parameters $\mu_i$ for the initial data 
can only be obtained numerically 
by solving the system of quadratic equations 
\eref{eq:ADM-Mass-general}. This takes by far most of 
the computational effort, so that we have to limit 
the number of black holes to about $10^5$.

First, we consider the configurations obtained from 
the pentatope-based Apollonian packings shown in 
\fref{fig:ApollonianPacking}. We calculate the 
different possibilities for the fitted radius:

\goodbreak

\begin{figure}[htb]
\centering
\includegraphics[width=0.7\columnwidth]{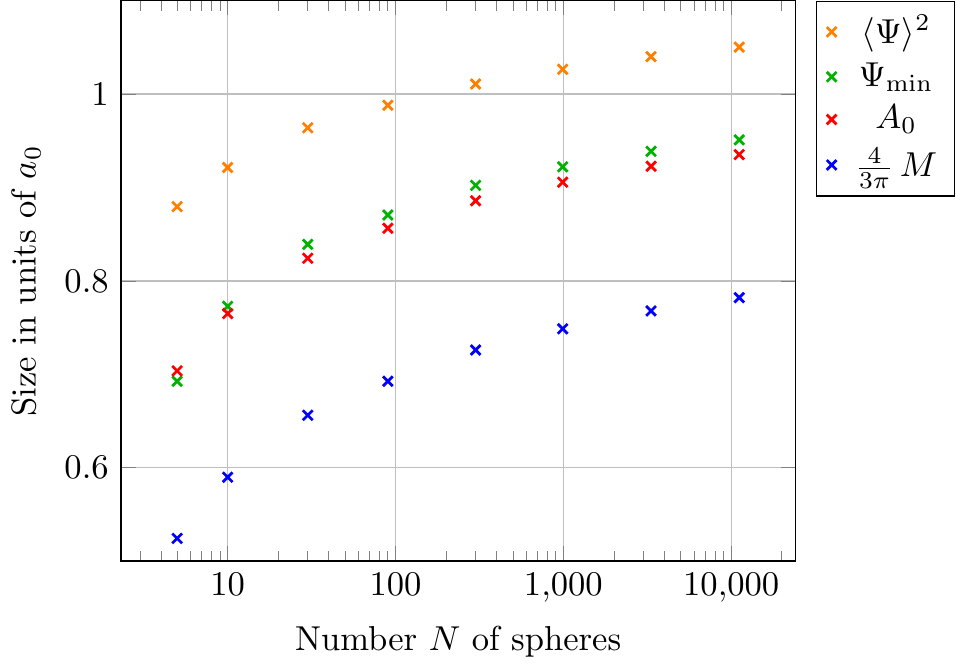}
\caption{Comparison between the different fits for 
the initial data with a swiss-cheese universe with size 
$a_0$. In both cases, the configuration is given by 
the Apollonian packing discussed above such that the 
black holes are located at the same positions with 
the same masses.}
\label{fig:Comparison1}
\end{figure}

\begin{enumerate}
\item our suggestion $ A_0 $ from \eref{eqn:UnifoamySize} 
\item for unifoamy configurations, \label{num:Size1}
\item Korzy\'nski's averaged value 
\item $ \langle \Psi \rangle^2 $, \label{num:Size2}
\item $ \frac{4}{3 \pi} M $ obtained from the total 
\item mass, \label{num:Size3}
\item the squared minimum of the conformal 
\item $\Psi_{\mathrm{min}}^2$. \label{num:Size4}
\end{enumerate}
The results for the first eight iterations of the 
pentatope-based Apollonian configurations are 
shown in \fref{fig:Comparison1}. 
All values are given in units of the size $a_0$ of the 
swiss-cheese dust universe. Hence, the best fit should 
approach the value $1$. However, we observe that the 
values differ from each other substantially and none 
really approaches the dust universe size; although 
the unifoamy size \eref{num:Size1} and the squared 
minimum \eref{num:Size2} seem to approach this 
value, they actually miss it. Furthermore, our 
suggestion \eref{num:Size1} differs strongly from 
the averaged value \eref{num:Size2} but it is closer 
to the squared minimum. Note that Korzy\'nski's 
first theorem would give almost the same (large) 
upper bounds for the deviation from the minimum 
because the spherical cap discrepancy should not 
really differ for the different iterations because 
we keep the biggest caps. For unifoamy configurations, 
our suggestion should be close to the size 
\eref{num:Size4} derived from the total mass, but this 
is not the case. Actually, if we check the unifoamy 
conditions \eref{eqn:UnifoamyMass} for all masses, we 
notice that they are violated by the biggest masses. 
Besides the spherical cap discrepancy, this indicates 
that very big masses are not possible for Friedmann-like 
configurations. This is consistent with our expectation 
that the masses in Friedmann-like configurations should 
be distributed somehow uniformly. In the considered 
configurations, the five biggest masses contained 
about half of the total mass.

\begin{figure}[htb]
\centering
\includegraphics[width=0.7\columnwidth]{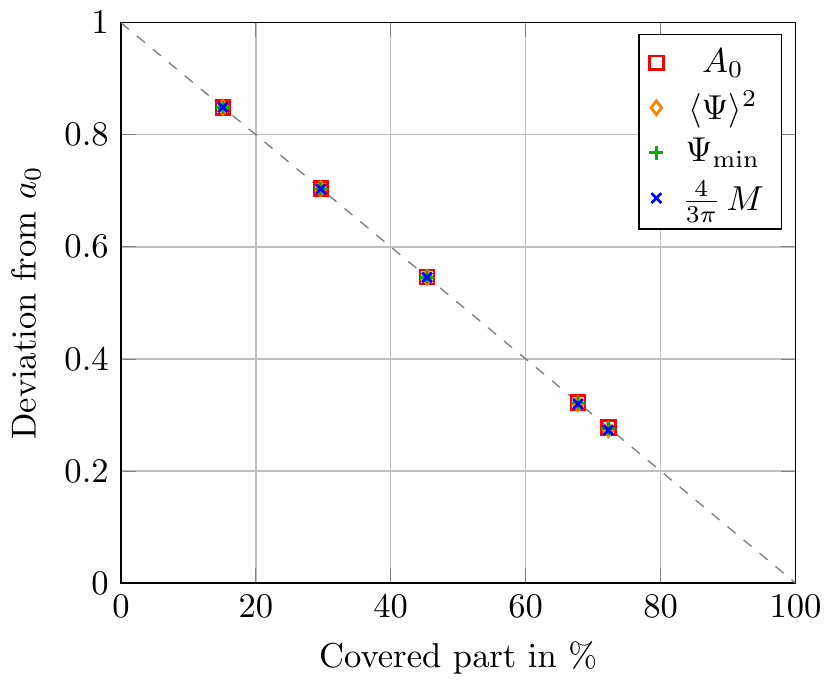}
\caption{Modified Apollonian configurations such that 
the biggest masses are replaced by smaller black holes. 
All five configurations contain about 12\,000 black holes.}
	\label{fig:Comparison2}
\end{figure}

In order to achieve a more uniform configuration, we 
substitute the biggest spheres by smaller ones by using 
the method described above. In \fref{fig:Comparison2}, 
we have plotted the deviation from the size of the 
swiss-cheese dust universe for five new configurations 
obtained from the pentatope-based Apollonian configuration 
with different maximal sizes for the spherical caps. 
In order to reduce the computational effort, we have also 
removed the smallest caps so that we have about $10^5$ 
masses in all five cases. This time, the configurations 
are approximately unifoamy and therefore the different 
results are in good agreement with each other. However, 
we also observe that the deviation from the swiss-cheese 
value is often quite large. But if we check how 
much of the dust universe in the swiss-cheese model 
is removed, we observe that the fit becomes better 
the less dust is remaining, which clearly fits expectation. 
In fact, the deviation appears to be proportional 
to the amount of remaining dust or, equivalently, 
the volume to the part of the 3-sphere that is 
uncovered by spherical caps. We conclude that the 
unifoamity of a configuration is not sufficient to 
guarantee a good fit, we also need an effective 
covering of the 3-sphere in the sense just explained.

The number of black holes in our computations is 
mainly limited by numerical reasons in calculating 
the mass parameters. We mention that the 
mass parameters can be estimated by
\begin{equation}
	\mu_i \approx m_i \,\sqrt{\frac{3 \pi}{16 M}}
\end{equation}
for unifoamy configurations, so that this step may 
be skipped leading to small deviations between the 
masses of the swiss-model and the initial data. 
Furthermore, it is also possible to use the masses 
$m_i$ instead of the mass parameters $\mu_i$ in 
order to check if a given configuration is 
unifoamy. This is true because it can be shown 
that
\begin{equation}
	\frac{\mu_i}{\mu} \approx \frac{m_i}{M}.
\end{equation}

This ends our first small excursion into 
applications of Lie sphere-geometry to lattice 
cosmology. We hope to have convinced the reader
that this is not only a beautiful but also 
very powerful method for the systematic 
construction of black-hole configurations of
almost arbitrary degrees of symmetry. 
We regard this paper only as a first step in
this direction, the primary purpose of which 
is to introduce the method and explain its 
geometric foundations. We are convinced that 
a proper geometric understanding is essential
in order to bring this method to its full power. 
Further work will be devoted to more concrete applications. 
 
\ack

The authors acknowledge financial support from the 
Research Training Center ``Models of Gravity'' 
funded by the DFG (Deutsche Forschungsgemeinschaft). 

\appendix

\section{Solution of Lichnerowicz equation on $S^3$}
\label{sec:Appendix1}
In this appendix we give a simple and general 
argument that implies that \eref{eq:Solution-LichnerowiczEquation_2} solves 
\eref{eq:LichnerowiczEquation_2}. This fact is 
a special case of the following general
\medskip

\textbf{Theorem.}
Let $\Delta_{S^n}$ denote the laplacian on the unit 
$n$-sphere which we think of as being embedded into 
$(n+1)$ - dimensional euclidean space: $S^n:=\{\bi{x}\in\mathbb{R}^{n+1}:\Vert\bi{x}\Vert=1\}$. Let 
$\bi{E}$ denote an arbitrary element of $S^n$, 
locally parametrised by some $n$ coordinates, 
like generalised polar angles, and $\bi{P}\in S^n$
a fixed point. We define the strictly positive 
function $D:S^n-\{\bi{P}\}\rightarrow\mathbb{R}$, $D(\bi{E}):=\Vert\bi{E}-\bi{P}\Vert$ which associates 
with each $\bi{E}\in S^n-\{\bi{P}\}$ its distance 
to $\bi{P}$ along the straight in $\mathbb{R}^{n+1}$. 
In other words: $D(\bi{E})$ denotes the geodesic 
distance of $\bi{E}$ from $\bi{P}$ as measured in the 
embedding $\mathbb{R}^{n+1}$, \emph{not} the intrinsic 
geodesic distance in $S^n$ (which is obviously always 
strictly larger). Then the theorem states that 
$D^{-(n-2)}$ is an eigenfunction of the laplacian on 
$S^n-\bi{P}$ with eigenvalue $n(n-2)/4$:
\begin{equation}
\label{eq:Appendix-1}
\Delta_{S^n}D^{-(n-2)}=\frac{n(n-2)}{4}\cdot D^{-(n-2)}\,.
\end{equation}
In particular, for $n=3$ we get $\Delta_{S^3}D^{-1}
=\frac{3}{4}\cdot D^{-1}$, which is just the statement 
that \eref{eq:Solution-LichnerowiczEquation_2} solves 
\eref{eq:LichnerowiczEquation_2}.

\medskip
\textbf{Proof.} Consider the function 
$\tilde D:\mathbb{R}^{n+1}\rightarrow\mathbb{R}$,
$\tilde D(r\bi{E}):=\Vert r\bi{E}-\bi{P}\Vert$, 
where $r\bi{E}$ denotes a general point in 
$\mathbb{R}^{n+1}-\{0\}$ whose norm is just $r>0$. The 
function $\tilde{D}=$ just extends $D$, i.e.,
$\tilde{D}\big\vert_{S^n}=D$. Now, the laplacian on
$\mathbb{R}^{n+1}$ can be written as follows:
\begin{equation}
\label{eq:Appendix-2}
\Delta_{\mathbb{R}^{n+1}}=
\partial^2_r+\frac{n}{r}\partial_r+
r^{-2}\,\Delta_{S^n}\,.
\end{equation}
This formula allows us to calculate the laplacian
of any real-valued function $F$ on
(an open subset of) $S^n$ by means of the laplacian 
of any extension $\tilde F$ of it to 
$\mathbb{R}^{n+1}$ (which is much easier to compute) 
and further simple $r$-differentiations. The formula 
we are using is: 
\begin{equation}
\label{eq:Appendix-3}
\Delta_{S^n}F=
\left(
\Delta_{\mathbb{R}^{n+1}}
-\partial_r^2-\frac{n}{r}\partial_r
\right)\Big\vert_{r=1} \tilde F
\,.
\end{equation}

In our case we have $\tilde{D}(\bi{rE})=(r^2-2rf+1)^{1/2}$,
where $f:=\bi{E}\cdot\bi{P}$ is a real valued 
function on $S^n$, independent of $r$. Simple 
calculations now show that 
\begin{equation}
\label{eq:Appendix-4}
\tilde{D}_1=D=\sqrt{2(1-f)}\,,\quad
\tilde{D}'_1=\frac{1}{2}D\,,\quad
\tilde{D}''_1=-\frac{1}{4}D+\frac{1}{D}\,,	
\end{equation}
where a prime denotes differentiation with respect to 
$r$ and the subscript $1$ indicates the restriction 
of the respective function (after differentiation) 
to $S^n$, i.e. $r=1$. 

Now we take $\tilde F={\tilde D}^{-k}$. The laplacian
of that in $\mathbb{R}^{n+1}$ is very easy to calculate,
e.g., by using spherical polar coordinates based at 
$\bi{P}$, in which case, using $\rho$ as radial 
coordinate, we have $\tilde D(\bi{E})=\rho$ and 
$\Delta_{\mathbb{R}^{n+1}}=\partial^2_\rho+(n/\rho)\partial_\rho$,
so that 
\begin{equation}
\label{eq:Appendix-5}
\Delta_{\mathbb{R}^{n+1}}\Big\vert_{r=1}\tilde{D}^{-k}
=k(k+1-n)\,D^{-k-2}\,.
\end{equation}
Furthermore, using \eref{eq:Appendix-4} a short 
computation shows 
\begin{equation}
\label{eq:Appendix-6}
\bigl(\partial^2_r+(n/r)\partial_r\bigr)
\Big\vert_{r=1}\tilde{D}^{-k}=
-k\,D^{-k-2}+\frac{k}{4}\bigl(k-2n+2\bigr)\,D^{-k}\,.
\end{equation}
Hence \eref{eq:Appendix-3} applied to $F=D^{-k}$ gives
\begin{equation}
\label{eq:Appendix-7}
\Delta_{S^n}D^{-k}=
k(k+2-n)\,D^{-k-2}+\frac{k}{4}\bigl(2n-k-2\bigr)\,D^{-k}\,.
\end{equation}
If we choose $k=n-2$ the first term vanishes and 
$D^{-k}=D^{2-n}$ becomes an (unbounded) eigenfunction 
of $\Delta_{S^n}$ on $S^n-\{\bi{P}\}$ with eigenvalue 
$n(n-2)/4$, as stated in \eref{eq:Appendix-1}.

\section{Stereographic projection and its metric properties}
\label{sec:Appendix2}
In this appendix we recall some properties of the 
stereographic projection from the unit $n$-sphere in 
$\mathbb{R}^{(n+1)}$ (or any euclidean vector space
of that dimension) onto its equatorial plane and 
the relation between the euclidean distances of 
source- and image points.

We consider $\mathbb{R}^{(n+1)}$ with the usual 
euclidean inner product and norm. As before, the 
latter will be denoted by $\Vert\cdot\Vert$. Again we 
consider the embedded unit $n$-sphere  
$S^n:=\{\bi{X}\in\mathbb{R}^{(n+1)} : \Vert X\Vert=1\}$. 
Points in $\mathbb{R}^{(n+1)}$ which lie on $S^n$ 
are denoted by capital bold-faced letters, like 
$\bi{X,Y}$, etc. Their inner product, according to the euclidean structure in $\mathbb{R}^{(n+1)}$, will be 
denoted by a dot, like $\bi{X}\cdot\bi{Y}$; hence, e.g., 
$\bi{X}^2:=\bi{X}\cdot\bi{X}=\Vert\bi{X}\Vert^2$.

We select a point $\bi{P}\in S^3$, called the 
``pole'', which will serve us as centre of 
the stereographic projection. Further, we let 
$\bi{P}^\perp:=\{\bi{X}\in\mathbb{R}^{(n+1)} : 
\bi{X}\cdot\bi{P}=0\}\simeq\mathbb{R}^n$ 
be the ``equatorial plane'' (a linear subspace),
elements of which we denote by lower case bold-faced 
letters,  like $\bi{x,y}$. The subspace $\bi{P}^\perp$ 
inherits a euclidean structure and norm from 
$\mathbb{R}^{(n+1)}$, which we continue to denote 
by a dot and $\Vert\cdot\Vert$, respectively.  

The given data define a diffeomorphism 
$\pi:S^n-\{\bi{P}\}\rightarrow\bi{P}^\perp$. 
It is called the stereographic projection from the pole 
onto the equatorial plane and is given by assigning to 
any $\bi{X}\in S^n-\{\bi{P}\}$ the unique 
intersection point of the line through $\bi{X}$ and 
$\bi{P}$ with $\bi{P}^\perp$. The parametric form 
(parameter $\lambda\in\mathbb{R}$)of the line is 
given by 
$\bi{L}(\lambda)=\bi{S}+\lambda (\bi{X}-\bi{P})$ 
and its intersection with $\bi{P}^\perp$ by  
$\bi{L}(\lambda_*)$, where $\lambda_*$ follows from  
$\bi{L}(\lambda_*)\cdot\bi{P}=0$. This gives 
\begin{equation}
\label{eq:Appendix2-1}
\bi{x}:=\pi(\bi{X})=\frac{\bi{X}-\bi{P}\,(\bi{P}\cdot\bi{X})}{1-\bi{P}\cdot\bi{X}}\,.	
\end{equation}
Its inverse is given by 
\begin{equation}
\label{eq:Appendix2-2}
\bi{X}=\pi^{-1}(\bi{x})=
\bi{x}\,\frac{2}{\bi{x}^2+1}
+\bi{P}\,\frac{\bi{x}^2-1}{\bi{x}^2+1}\,.
\end{equation}
Equations \eref{eq:Appendix2-1} and 
\eref{eq:Appendix2-2} define the stereographic 
diffeomorphism between the once-punctured $n$-sphere
and the equatorial $n$-plane. 

Next we wish to relate the euclidean distances between 
source- and image points. We start by noting that 
\begin{equation}
\label{eq:Appendix2-3}
\Vert\bi{X}-\bi{P}\Vert^2=2(1-\bi{X}\cdot\bi{P})=\frac{4}{1+\bi{x}^2}\,,
\end{equation}
where we used $\bi{X}^2=\bi{P}^2=1$ and \eref{eq:Appendix2-2} 
with $\bi{x}\cdot\bi{P}=0$ in the 2nd step. 
Similarly, for $\bi{X}:=\pi^{-1}(\bi{x})$ and 
$\bi{Y}:=\pi^{-1}(\bi{y})$, equation 
\eref{eq:Appendix2-2} yields
\begin{equation}
\label{eq:Appendix2-4}
\bi{X}\cdot\bi{Y}=
\frac{4\,\bi{x}\cdot\bi{y}+(\bi{x}^2-1)(\bi{y}^2-1)}{(\bi{x}^2+1)(\bi{y}^2+1)}\,,
\end{equation}
and hence
\begin{eqnarray}
\label{eq:Appendix2-5}
4\,\Vert\bi{X}-\bi{Y}\Vert^2
&=8(1-\bi{X}\cdot\bi{Y})=
\frac{16(\bi{x}-\bi{y})^2}
{(1+\bi{x}^2)(1+\bi{y}^2)}\nonumber\\
&=\Vert\bi{x}-\bi{y}\Vert^2\
\Vert\bi{X}-\bi{P}\Vert^2\
\Vert\bi{Y}-\bi{P}\Vert^2\,,
\end{eqnarray}
using \eref{eq:Appendix2-3} for $\bi{X}$
and $\bi{Y}$ in the last step. This leads to 
the final relation
\begin{equation}
\label{eq:Appendix2-6}
\Vert\bi{x}-\bi{y}\Vert=\frac{2\,\Vert\bi{X}-\bi{Y}\Vert}%
{\Vert\bi{X}-\bi{P}\Vert\
\Vert\bi{Y}-\bi{P}\Vert}\,.
\end{equation}
that holds independently of the dimensions $n$ 
and that we used in \eref{eq:BrillLindquist-Isom3}.	

The Riemannian metric of $S^n$ is that induced by the 
embedding $S^n\hookrightarrow\mathbb{R}^{(n+1)}$. In
stereographic coordinates $\bi{x}\in\bi{P}^\perp$ this
metric follows from pulling back the Riemannian metric
on $S^n$ via the inverse stereographic projection $\pi^{-1}$.
This is easily computed from \eref{eq:Appendix2-2} by 
first calculating the differential of $\bi{X}(\bi{x})$,
\begin{equation}
\label{eq:Appendix2-7}
\mathbf{d}\bi{X}
=\frac{2}{1+\bi{x}^2}\,\mathbf{d}\bi{x}
+\frac{4\,(\bi{P}-\bi{x})}{(\bi{x}^2+1)^2}\,(\bi{x}\cdot \mathbf{d}\bi{x})
\end{equation} 
and then `squaring' it, $d\bi{X}\dot\otimes d\bi{X}:=\delta_{ab}dX^a\otimes dX^b$, which immediately gives, taking into 
account $\bi{x}\cdot\bi{P}=0$ and  $d\bi{x}\cdot\bi{P}=0$,
\begin{equation}
\label{eq:Appendix2-8}
\mathbf{d}\bi{X}\dot\otimes\mathbf{d}\bi{X}=
=\left(\frac{2}{1+\bi{x}^2}\right)^2\,
\mathbf{d}\bi{x}\dot\otimes \mathbf{d}\bi{x}\,.
\end{equation} 
Comparison with \eref{eq:Appendix2-3} shows 
that the flat metric $\bi{h}_{\mathbb{R}^n}:=d\bi{x}\cdot d\bi{x}$ on $\bi{P}^\perp\cong\mathbb{R}^n$ can 
be written in terms of the constant positive-curvature metric on the unit n-sphere,
$\bi{h}_{S^n}:=(\pi^{-1})^*(\mathbf{d}\bi{X}\cdot \mathbf{d}\bi{X})$, as follows: 
\begin{equation}
\label{eq:Appendix2-9}
\bi{h}_{\mathbb{R}^n}=\frac{4}{\Vert\bi{X}-\bi{P}\Vert^4}\,\bi{h}_{S^n}\,.
\end{equation} 
This is the equation we used in 
\eref{eq:BrillLindquist-Isom1}.

\end{document}